\documentclass[aps,prd,preprint,groupedaddress]{revtex4}

\usepackage{graphicx}
\usepackage{side}
\usepackage{axodraw}

\def\dfrac#1#2{\displaystyle\frac{#1}{#2}}

\newcommand{\p}{\partial}
\newcommand{\kslash}{k\kern-1ex /}
\newcommand{\pslash}{p\kern-1ex /}
\newcommand{\qslash}{q\kern-1ex /}
\newcommand{\lslash}{l\kern-1ex /}
\newcommand{\sslash}{s\kern-1ex /}
\newcommand{\Dslash}{{\cal D}\kern-1.5ex /}

\newcommand{\tr}{{\rm tr}}

\newcommand{\beqa}{\begin{eqnarray}}
\newcommand{\eeqa}{\end{eqnarray}}

\newcommand{\be}{\begin{equation}}
\newcommand{\ee}{\end{equation}}
\newcommand{\ben}{\begin{eqnarray}}
\newcommand{\een}{\end{eqnarray}}
\newcommand{\nn}{\nonumber}
\def\lsim{\raise0.3ex\hbox{$<$\kern-0.75em\raise-1.1ex\hbox{$\sim$}}}
\def\gsim{\raise0.3ex\hbox{$>$\kern-0.75em\raise-1.1ex\hbox{$\sim$}}}
\def\simgt{\rlap{\lower 3.5 pt\hbox{$\mathchar \sim$}}\raise 1pt \hbox {$>$}}
\def\simlt{\rlap{\lower 3.5 pt\hbox{$\mathchar \sim$}}\raise 1pt \hbox {$<$}}
\newcommand{\cont}{{\rm cont}}
\newcommand{\latt}{{\rm latt}}

\newcommand{\mf}{{\rm MF}}
\newcommand{\msbar}{{\overline {\rm MS}}}

\newcommand{\csw}{{c_{\rm SW}}}
\newcommand{\hk}{{\hat k}}

\newcommand{\ce}{{\it c}_{\it E}}
\newcommand{\cb}{{\it c}_{\it B}}
\newcommand{\lqcd}{{\Lambda}_{\rm QCD}}
\newcommand{\pos}{{p^*}}
\newcommand{\qos}{{q^*}}
\newcommand{\lo}{{(0)}}
\newcommand{\nlo}{{(1)}}
\newcommand{\mplo}{{m_p^{(0)}}}
\newcommand{\intlat}{{\int_{-\pi}^{\pi}\frac{d^4 k}{(2\pi)^4}}}
\newcommand{\mplomf}{{{\tilde m_p}^{(0)}}}

\begin{document}


\title{A Perturbative Determination of \\
Mass Dependent $O(a)$ Improvement Coefficients \\
in a Relativistic Heavy Quark Action}


\author{Sinya Aoki$^a$, Yasuhisa Kayaba$^a$ and Yoshinobu Kuramashi$^b$}
\affiliation{$^a$Institute of Physics, University of Tsukuba, 
Tsukuba, Ibaraki 305-8571, Japan \\
$^b$Institute of Particle and Nuclear Studies,
High Energy Accelerator Research Organization(KEK),
Tsukuba, Ibaraki 305-0801, Japan}


\date{\today}

\begin{abstract}
We present the results for a perturbative determination 
of mass dependent improvement coefficients $\nu$, $r_s$, $c_E$ and $c_B$ 
in a relativistic heavy quark action, which 
we have designed to control $m_Qa$ errors by extending the 
on-shell  $O(a)$ improvement program to 
the case of $m_Q \gg \Lambda_{\rm QCD}$, where $m_Q$ is the heavy quark mass.
The parameters $\nu$ and $r_s$ are determined from the quark propagator
and $c_E$ and $c_B$ are from the on-shell 
quark-quark scattering amplitude.
We show that all the parameters, together with the quark wave function 
and the mass renormalization factors, are determined free from infrared 
divergences once their tree level values are properly tuned.
Results of these parameters are shown as a function of $m_Q a$ 
for various improved gauge actions.

\end{abstract}


\maketitle


\section{Introduction}
\label{sec:intro}

A calculation of weak matrix elements for the $B$ and $D$ mesons 
is a subject of great interest in lattice QCD: their precise determination
is an essential ingredient to extract the Cabibbo-Kobayashi-Maskawa
matrix. Although in principle we can extract these weak matrix elements
precisely from lattice QCD simulations, 
it is still difficult to achieve this goal.
A main obstacle is the systematic error originating from
large $m_Q a$ corrections 
with current accessible computational resources: 
$m_b a\sim 1-2$ and $m_c a\sim 0.3-0.6$ in 
the quenched approximation and $m_b a\sim 2-3$ and $m_c a\sim 0.6-0.9$
in unquenched QCD.

To control this large $m_Q a$ corrections 
a new relativistic approach is
proposed from the view point of the on-shell $O(a)$
improvement program\cite{akt}. 
The generic quark action is given by
\ben
S_q&=&\sum_x\left[ m_0{\bar q}(x)q(x)
+{\bar q}(x)\gamma_0 D_0q(x)
+\nu \sum_i {\bar q}(x)\gamma_i D_i q(x)\right.\nn\\
&&\left.-\frac{r_t a}{2} {\bar q}(x)D_0^2 q(x)
-\frac{r_s a}{2} \sum_i {\bar q}(x)D_i^2 q(x)\right.\nn\\
&&\left.-\frac{ig a}{2}c_E \sum_i {\bar q}(x)\sigma_{0i}F_{0i} q(x)
-\frac{ig a}{4}c_B \sum_{i,j} {\bar q}(x)\sigma_{ij}F_{ij} q(x)
\right],
\een
where we are allowed to choose $r_t=1$ and other four parameters
$\nu$, $r_s$, $c_E$ and $c_B$ are analytic functions of $m_Q a$ and 
the gauge coupling constant $g$.

In this formulation
the leading cutoff effects of order $(m_Qa)^n$ are absorbed
in the definition of renormalization factors for the quark mass 
and the wave function. 
After removing the next-leading cutoff effects of
$O((m_Qa)^n a\lqcd)$ with $\nu$, $r_s$, $c_E$ and $c_B$  in the quark action 
properly adjusted in the $m_Qa$ dependent way, we are left with at most
$O((a\lqcd)^2)$ errors.

In this paper we calculate $\nu$, $r_s$, $c_E$ and $c_B$ up to
the one-loop level for various improved gauge actions. While  
$\nu$ and $r_s$, together with the quark wave function renormalization factor
$Z_q$ and the pole mass $m_p$ as byproducts, are obtained from the quark 
propagator, 
$c_E$ and $c_B$ are determined from the on-shell 
quark-quark scattering amplitude.
One-loop diagrams are evaluated 
by employing the conventional perturbative method with the use of 
the fictitious gluon mass to regularize the infrared divergence,
which was successfully applied in the massless case
to the calculation of the renormalization constants and the
improvement coefficients for the quark bilinear operators\cite{gmass}
and the improvement coefficient $\csw$\cite{csw_m0}. 
We show that the improvement parameters $\nu$, $r_s$, $c_E$ and $c_B$
are determined free from the infrared divergences, 
once their tree level values are correctly tuned
in the $m_Q a$ dependent way.
Some preliminary results are presented in Ref.~\cite{latt02}.

This paper is organized as follows. In Sec.~II we introduce the 
relativistic heavy quark action and the improved gauge actions.
We also give their Feynman rules relevant for the
present calculation. We describe 
the method for the calculation of the one-loop diagrams in Sec.~III.
In Sec.~IV we determine $\nu$ and $r_s$ 
from the quark propagator. The results for $Z_q$ and $m_p$ 
are also presented. In Sec.~V $\cb$ and $\ce$ are determined
from the on-shell quark-quark scattering amplitude.
We give a rather detailed description on the cancellation of
the infrared divergences in one-loop diagrams.
In Sec.~VI we explain how to implement the mean field improvement
of the parameters.
Our conclusions are summarized in Sec.~VII. 

The physical quantities are expressed in lattice units and 
the lattice spacing $a$ is suppressed unless necessary.
We take  SU($N_c$) gauge group with the gauge coupling constant $g$.

\section{Action and Feynman Rules}
\label{sec:action}
 
The relativistic heavy quark action proposed in Ref.~\cite{akt}
is given by 
\ben
S_{\rm quark}&=&
\sum_n\frac{1}{2}
\left\{{\bar \psi}_n(-r_t+\gamma_0)U_{n,0}\psi_{n+{\hat 0}}
      +{\bar \psi}_n(-r_t-\gamma_0)U^\dagger_{n-{\hat 0},0}
       \psi_{n-{\hat 0}}\right\}\nn\\
&&+\sum_n\frac{1}{2}\sum_i
\left\{{\bar \psi}_n(-r_s+\nu\gamma_i)U_{n,i}\psi_{n+{\hat i}}
      +{\bar \psi}_n(-r_s-\nu\gamma_i)U^\dagger_{n-{\hat i},i}
       \psi_{n-{\hat i}}\right\}\nn\\
&&+(m_0+r_t+3r_s)\sum_n{\bar \psi}_n\psi_n \nn\\
&&-\ce\sum_n\sum_{i}ig\frac{1}{2}
{\bar \psi}_n\sigma_{0i}F_{0i}(n)\psi_n
-\cb\sum_n\sum_{i,j}ig\frac{1}{4}
{\bar \psi}_n\sigma_{ij}F_{ij}(n)\psi_n,
\label{eq:action_q}
\een
where we define the Euclidean gamma matrices in terms of
the Minkowski ones in the Bjorken-Drell convention:
$\gamma_0=\gamma_{BD}^0$,
$\gamma_j=-i\gamma_{BD}^j$ $(j=1,2,3)$, 
$\gamma_5=\gamma_{BD}^5$ and 
$\sigma_{\mu\nu}=\frac{1}{2}[\gamma_\mu,\gamma_\nu]$.
Whereas the value of $r_t$ can be chosen arbitrarily,
$\nu$, $r_s$, $\ce$ and $\cb$ have to be adjusted 
to remove the cutoff effects of
$O((m_Qa)^n a\lqcd)$. As explained in Ref.~\cite{akt} 
the $(m_Q a)^n$ corrections can be avoided by the redefinition
of the quark field and mass. 
The field strength $F_{\mu\nu}$ in the clover term
is expressed as
\ben 
F_{\mu\nu}(n)&=&\frac{1}{4}\sum_{i=1}^{4}\frac{1}{2ig}
\left(U_i(n)-U_i^\dagger(n)\right), \\
U_1(n)&=&U_{n,\mu}U_{n+{\hat \mu},\nu}
         U^\dagger_{n+{\hat \nu},\mu}U^\dagger_{n,\nu}, \\
U_2(n)&=&U_{n,\nu}U^\dagger_{n-{\hat \mu}+{\hat \nu},\mu}
         U^\dagger_{n-{\hat \mu},\nu}U_{n-{\hat \mu},\mu}, \\
U_3(n)&=&U^\dagger_{n-{\hat \mu},\mu}U^\dagger_{n-{\hat \mu}-{\hat \nu},\nu}
         U_{n-{\hat \mu}-{\hat \nu},\mu}U_{n-{\hat \nu},\nu}, \\
U_4(n)&=&U^\dagger_{n-{\hat \nu},\nu}U_{n-{\hat \nu},\mu}
         U_{n+{\hat \mu}-{\hat \nu},\nu}U^\dagger_{n,\mu}.
\een
The weak coupling perturbation theory is developed  
by writing the link variable in terms of the gauge potential
\be
U_{n,\mu}=\exp \left(igaT^A A_\mu^A\left(n+\frac{1}{2}\hat\mu\right)\right),
\ee
where $T^A$ ($A=1,\dots,N^2_c-1$) is a generator of color SU($N_c$).

The quark propagator is obtained by inverting the Wilson-Dirac operator 
in eq.(\ref{eq:action_q}),
\ben
S_q^{-1}(p)&=&i \gamma_0 {\rm sin}(p_0)
+\nu i \sum_i \gamma_i {\rm sin}(p_i) +m_0 \nn\\
&&+r_t (1-{\rm cos}(p_0))+r_s\sum_i(1-{\rm cos}(p_i)),
\label{eq:qprop}
\een
For the present calculation, we need 
one-, two- and three-gluon vertices with quarks:
\ben
V^A_{10}(p,q)&=&-gT^A\left\{
i\gamma_0{\rm cos}\left(\frac{p_0+q_0}{2}\right)
+r_t{\rm sin}\left(\frac{p_0+q_0}{2}\right)\right\}, 
\label{eq:vtx_w1t}\\
V^A_{1i}(p,q)&=&-gT^A\left\{
\nu i\gamma_i{\rm cos}\left(\frac{p_i+q_i}{2}\right)
+r_s{\rm sin}\left(\frac{p_i+q_i}{2}\right)\right\}, 
\label{eq:vtx_w1s}\\
V_{200}^{AB} (p,q)
&=& \frac{a}{2} g^2 \frac{1}{2} \{T^{A}, T^{B}\}
\left\{ i \gamma_0 \sin \left(\frac{p_0+q_0}{2}\right)
-r_t \cos \left(\frac{p_0+q_0}{2}\right)\right\},
\label{eq:vtx_w2t}\\
V_{2ii}^{AB} (p,q)
&=& \frac{a}{2} g^2 \frac{1}{2} \{T^{A}, T^{B}\}
\left\{\nu i \gamma_i \sin \left(\frac{p_i+q_i}{2}\right)
-r_s \cos \left(\frac{p_i+q_i}{2}\right)\right\},
\label{eq:vtx_w2s}\\
V_{300}^{ABC} (p,q)
&=& \frac{a^2}{6} g^3 \frac{1}{6} 
\left[T^{A}\{T^{B},T^{C}\}+T^{B}\{T^{C},T^{A}\}+T^{C}\{T^{A},T^{B}\}\right]
\nn\\
&&\times\left\{i\gamma_0 \cos \left(\frac{p_0+q_0}{2}\right)
+r_t \sin \left(\frac{p_0+q_0}{2}\right)\right\},
\label{eq:vtx_w3t}\\
V_{3iii}^{ABC} (p,q)
&=& \frac{a^2}{6} g^3 \frac{1}{6} 
\left[T^{A}\{T^{B},T^{C}\}+T^{B}\{T^{C},T^{A}\}+T^{C}\{T^{A},T^{B}\}\right]
\nn\\
&&\times\left\{\nu i\gamma_i \cos \left(\frac{p_i+q_i}{2}\right)
+r_s \sin \left(\frac{p_i+q_i}{2}\right)\right\},
\label{eq:vtx_w3s}\\
V^A_{c1\mu}(p,q)&=&-gT^A\frac{1}{2}
\sum_\nu \csw^{\mu\nu}\sigma_{\mu\nu}
\cos \left(\frac{p_\mu-q_\mu}{2}\right)\sin (p_\nu-q_\nu),
\label{eq:vertex_c1}\\
V^{AB}_{c2\mu\nu}(p,q,k_1,k_2)&=&
-\frac{a}{2} g^2 if_{ABC}T^C\frac{1}{4}\nn\\
&&\times\left\{
\csw^{\mu\nu}\sigma_{\mu\nu}\left[
4\cos \left(\frac{k_{1\nu}}{2}\right)
\cos \left(\frac{k_{2\mu}}{2}\right)
\cos \left(\frac{q_\mu-p_\mu}{2}\right)
\cos \left(\frac{q_\nu-p_\nu}{2}\right)\right.\right.\nn\\
&&\left.\left.
-2\cos \left(\frac{k_{1\mu}}{2}\right)
\cos \left(\frac{k_{2\nu}}{2}\right)\right]\right.\\
&&\left.+\delta_{\mu\nu}\sum_\rho\csw^{\mu\rho}\sigma_{\mu\rho}
\sin \left(\frac{q_\mu-p_\mu}{2}\right)
\left[\sin(k_{2\rho})-\sin(k_{1\rho})\right]\right\},\nn
\label{eq:vertex_c2}\\
V^{ABC}_{c3\mu\nu\tau}(p,q,k_1,k_2,k_3)
&=&-3ig^3 \frac{a^2}{6}\nn\\
&&\times\left[ T^{A}T^{B}T^{C}
\delta_{\mu\nu}\delta_{\mu\tau}
\sum_\rho i\csw^{\mu\rho}\sigma_{\mu\rho}\left\{ 
-\frac{1}{6}\cos \left(\frac{q_\mu-p_\mu}{2}\right)
\sin (q_\rho-p_\rho)\right.\right.\nn\\
&&\left.\left.
+\cos \left(\frac{q_\mu-p_\mu}{2}\right)
\cos \left(\frac{q_\rho-p_\rho}{2}\right)
\cos \left(\frac{k_{3\rho}-k_{1\rho}}{2}\right)
\sin \left(\frac{k_{2\rho}}{2}\right)
\right\}\right.\nn\\
&&\left.
-\frac{1}{2}\left[T^A T^B T^C+T^C T^B T^A\right]
i\csw^{\mu\nu}\sigma_{\mu\nu}\right.\\
&&\left.
\times\left\{\delta_{\nu\tau}
2\cos \left(\frac{q_\mu-p_\mu}{2}\right)
\cos \left(\frac{q_\nu-p_\nu}{2}\right)
\cos \left(\frac{k_{3\mu}+k_{2\mu}}{2}\right)
\sin \left(\frac{k_{1\nu}}{2}\right)\right.\right.\nn\\
&&\left.\left.
+\delta_{\nu\tau}
\sin \left(\frac{k_{3\nu}+k_{2\nu}}{2}\right)
\cos \left(\frac{k_{1\mu}}{2}+k_{2\mu}\right)\right.\right.\nn\\
&&\left.\left.
+\delta_{\mu\tau}
\sin \left(\frac{k_{1\mu}+2k_{2\mu}+k_{3\mu}}{2}\right)
\cos \left(\frac{q_\nu-p_\nu}{2}\right)
\cos \left(\frac{k_{3\nu}-k_{1\nu}}{2}\right)
\right\}\right],\nn
\label{eq:vertex_c3}
\een
where $\csw^{0i}=\csw^{i0}=\ce$, $\csw^{ij}=\cb$ $(i,j=1,2,3)$ and 
$f_{ABC}$ is the structure constant of SU($N_c$) gauge group.
The first six vertices originate from 
the Wilson quark action and the last three from the clover term. 
The momentum assignments for the vertices are depicted 
in Fig.~\ref{fig:fr_vtx_qg}.

For the gauge  action we consider the following
general form including the standard plaquette term 
and six-link loop terms:
\be
S_{\rm g}=\frac{1}{g^2}\left\{ c_0\sum_{\rm plaquette}\tr U_{pl}
+c_1\sum_{\rm rectangle}\tr U_{rtg}
+c_2\sum_{\rm chair}\tr U_{chr}
+c_3\sum_{\rm parallelogram} \tr U_{plg} \right\}
\label{eq:action_g}
\ee
with the normalization condition
\be
c_0+8c_1+16c_2+8c_3=1,
\ee
where six-link loops are composed of a $1\times 2$ rectangle,
a bent $1\times 2$ rectangle (chair) and a three-dimensional
parallelogram.
In this paper we consider the following choices:
$c_1=c_2=c_3=0$(Plaquette),
$c_1=-1/12$, $c_2=c_3=0$(Symanzik)\cite{Weisz83,LW} 
$c_1=-0.331$, $c_2=c_3=0$(Iwasaki), $c_1=-0.27, 
c_2+c_3=-0.04$(Iwasaki')
\cite{Iwasaki83}, $c_1=-0.252, c_2+c_3=-0.17$(Wilson)\cite{Wilson}
and $c_1 = -1.40686$, 
$c_2=c_3=0$(doubly blocked Wilson 2 (DBW2))\cite{dbw2}.
The last four cases are called the RG improved gauge action whose 
parameters are chosen to be the values 
suggested by approximate renormalization group analyses.
Some of these actions are now getting widely used, since
they realize continuum-like gauge field fluctuations  
better than the naive plaquette action at the same lattice spacing. 

The free gluon propagator is derived in Ref.~\cite{Weisz83}:
\ben
D_{\mu\nu}(k)&=&\frac{1}{(\hk^2)^2}\left[
(1-A_{\mu\nu})\hk_\mu\hk_\nu+
\delta_{\mu\nu}\sum_\sigma\hk_\sigma^2 A_{\nu\sigma}\right]
\een
with 
\ben
\hk_\mu&=&2{\rm sin}\left(\frac{k_\mu}{2}\right),\qquad
\hk^2=\sum_{\mu=0}^{3}\hk_\mu^2.
\een
The matrix $A_{\mu\nu}$ satisfies
\ben
&({\rm i})& A_{\mu\mu}=0\;\;\; {\rm for}\;\;{\rm all}\;\; \mu, \\
&({\rm ii})& A_{\mu\nu}=A_{\nu\mu}, \\
&({\rm iii})& A_{\mu\nu}(k)=A_{\mu\nu}(-k). \\
&({\rm iv})& A_{\mu\nu}(0)=1\;\;\; {\rm for}\;\; \mu\ne\nu,
\een
and its expression is given by
\ben
A_{\mu\nu}(k)&=&\frac{1}{\Delta_4}
\left[(\hk^2-\hk_\nu^2)(q_{\mu\rho}q_{\mu\tau}\hk_\mu^2
+q_{\mu\rho}q_{\rho\tau}\hk_\rho^2
+q_{\mu\tau}q_{\rho\tau}\hk_\tau^2)\right. \nn\\
&&\left.+(\hk^2-\hk_\mu^2)(q_{\nu\rho}q_{\nu\tau}\hk_\nu^2
+q_{\nu\rho}q_{\rho\tau}\hk_\rho^2
+q_{\nu\tau}q_{\rho\tau}\hk_\tau^2)\right. \nn\\
&&\left.+q_{\mu\rho}q_{\nu\tau}(\hk_\mu^2+\hk_\rho^2)(\hk_\nu^2+\hk_\tau^2) 
+q_{\mu\tau}q_{\nu\rho}(\hk_\mu^2+\hk_\tau^2)(\hk_\nu^2+\hk_\rho^2)\right. \nn\\
&&\left.-q_{\mu\nu}q_{\rho\tau}(\hk_\rho^2+\hk_\tau^2)^2
-(q_{\mu\rho}q_{\nu\rho}+q_{\mu\tau}q_{\nu\tau})\hk_\rho^2\hk_\tau^2\right. \nn\\
&&\left.-q_{\mu\nu}(q_{\mu\rho}\hk_\mu^2\hk_\tau^2
           +q_{\mu\tau}\hk_\mu^2\hk_\rho^2
           +q_{\nu\rho}\hk_\nu^2\hk_\tau^2
           +q_{\nu\tau}\hk_\nu^2\hk_\rho^2)\right],
\label{eq:matrix_a}
\een
with $\mu\ne\nu\ne\rho\ne\tau$ the Lorentz indices. 
$q_{\mu\nu}$ and $\Delta_4$  are written as
\ben
q_{\mu\nu}&=&(1-\delta_{\mu\nu})
\left[1-(c_1-c_2-c_3)(\hk_\mu^2+\hk_\nu^2)-(c_2+c_3)\hk^2\right], \\
\Delta_4&=&\sum_\mu \hk_\mu^4\prod_{\nu\ne\mu}q_{\nu\mu}
+\sum_{\mu >\nu,\rho >\tau,\{\rho,\tau\}\cap\{\mu,\nu\}=\emptyset}
\hk_\mu^2\hk_\nu^2q_{\mu\nu}(q_{\mu\rho}q_{\nu\tau}
                            +q_{\mu\tau}q_{\nu\rho}).
\een
In the case of the standard plaquette action,
the matrix $A_{\mu\nu}$ is simplified as
\be
A_{\mu\nu}^{\rm plaquette}=1-\delta_{\mu\nu}.
\ee

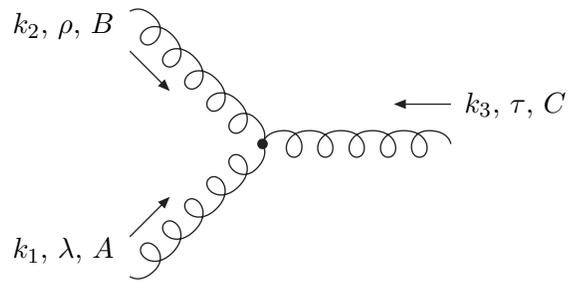
\begin{figure}[t]
\begin{center}

\begin{picture}(200,150)(0,50)
\Gluon(50,50)(100,100){-5}{5}
\Gluon(100,100)(50,150){-5}{5}
\Vertex(100,100){2}
\Gluon(100,100)(171,100){5}{5}
\LongArrow(50,65)(64,79)
\Text(25,60)[c]{$k_1$, $\lambda$, $A$}
\LongArrow(50,135)(64,121)
\Text(25,145)[c]{$k_2$, $\rho$, $B$}
\LongArrow(171,115)(151,115)
\Text(196,115)[c]{$k_3$, $\tau$, $C$}
\end{picture}

\end{center}
\caption{Momentum assignment for the three-gluon vertex.}
\label{fig:fr_vtx_3g}
\vspace{8mm}
\end{figure}

The present calculation requires only the three-point vertex which
is given in Ref.~\cite{Weisz83},
\be
V^{ABC}_{g3\lambda\rho\tau}(k_1,k_2,k_3)
=-i\frac{g}{6}f_{ABC}
\sum_{i=0}^{3}c_iV^{(i)}_{g3\lambda\rho\tau}(k_1,k_2,k_3)
\ee
with
\ben
V^{(0)}_{g3\lambda\rho\tau}(k_1,k_2,k_3)
&=&\delta_{\lambda\rho}\widehat{(k_1-k_2)}_\tau c_{3\lambda}
+{\rm 2\; cycl.\;\;perms.},\\
V^{(1)}_{g3\lambda\rho\tau}(k_1,k_2,k_3)
&=&8V^{(0)}_{g3\lambda\rho\tau}(k_1,k_2,k_3)\nn\\
&&+\left[ 
\delta_{\lambda\rho}\left\{ 
c_{3\lambda}(\widehat{(k_1-k_2)}_\lambda
(\delta_{\lambda\tau}\hk_3^2-\hk_{3\lambda}\hk_{3\tau})
-\widehat{(k_1-k_2)}_\tau(\hk_{1\tau}^2+\hk_{2\tau}^2))\right.
\right.
\nn\\
&&+
\left.
\left.\widehat{(k_1-k_2)}_\tau
(\hk_{1\lambda}\hk_{2\lambda}
-2c_{1\lambda}c_{2\lambda}\hk_{3\lambda}^2)\right\}+{\rm 2\; cycl.\;\;perms.}
\right]
,\\
V^{(2)}_{g3\lambda\rho\tau}(k_1,k_2,k_3)
&=&16V^{(0)}_{g3\lambda\rho\tau}(k_1,k_2,k_3)\nn\\
&&-\left[
\delta_{\lambda\rho}(1-\delta_{\lambda\tau})
c_{3\lambda}\sum_{\sigma\ne\lambda,\tau}
\left\{\widehat{(k_1-k_2)}_\tau
(\hk_{1\sigma}^2+\hk_{2\sigma}^2+\hk_{3\sigma}^2)
+\hk_{3\tau}(\hk_{1\sigma}^2-\hk_{2\sigma}^2)\right\}
\right.
\nn\\ 
&&\left.
+(1-\delta_{\lambda\rho})(1-\delta_{\lambda\tau})(1-\delta_{\rho\tau})
\hk_{1\lambda}\hk_{2\rho}\widehat{(k_1-k_2)}_\tau+{\rm 2\; cycl.\;\;perms.}
\right]
,\\
V^{(3)}_{g3\lambda\rho\tau}(k_1,k_2,k_3)
&=&8V^{(0)}_{g3\lambda\rho\tau}(k_1,k_2,k_3)\nn\\
&&-\left[
\delta_{\lambda\rho}(1-\delta_{\lambda\tau})
c_{3\lambda}\widehat{(k_1-k_2)}_\tau
\sum_{\sigma\ne\lambda,\tau}(\hk_{1\sigma}^2+\hk_{2\sigma}^2)
\right.
\nn\\
&&+\frac{1}{2}(1-\delta_{\lambda\rho})(1-\delta_{\lambda\tau})
(1-\delta_{\rho\tau})\widehat{(k_1-k_2)}_\tau
\left\{\hk_{1\lambda}\hk_{2\rho}
-\frac{1}{3}\widehat{(k_3-k_1)}_\rho\widehat{(k_2-k_3)}_\lambda \right\}\nn\\
&&\left. +{\rm 2\; cycl.\;\;perms.}
\right],
\een
where we introduce the notation,
\ben
c_{i\lambda}&=&\cos\left(\frac{k_{i\lambda}}{2}\right).
\een
The momentum assignments are found in Fig.~\ref{fig:fr_vtx_3g}.

\section{Method for the calculation of one-loop diagrams}
\label{sec:method}

We determine 
$Z_q=Z_q^\lo+g^2 Z_q^\nlo$, 
$m_p=m_p^\lo+g^2 m_p^\nlo$, 
$\nu=\nu^\lo+g^2 \nu^\nlo$,  
$r_s=r_s^\lo+g^2 r_s^\nlo$ from the quark propagator and
$\ce=\ce^\lo+g^2\ce^\nlo$, $\cb=\cb^\lo+g^2\cb^\nlo$ 
from the on-shell quark-quark
scattering amplitude, where the superscript $(i)$ denotes the
$i$-th loop level.
The tree level values of the parameters are already determined
in Ref.~\cite{akt}.
To calculate the one-loop contribution
we write a $Mathematica$ program for a given loop diagram 
to compose the integrand
of the Feynman rules and to practice the Dirac algebra. 
The output is then transformed into a FORTRAN code by $Mathematica$.
The momentum integration is performed by a mode sum for a periodic
box of a size $L^4$ with $L=64$ after transforming the momentum 
variable through $k^\prime_\mu=k_\mu-\sin k_\mu$.
The numerical errors are estimated by varying $L$ from 64 to 60.

Although $\nu^\nlo$, $r_s^\nlo$, $\ce^\nlo$ and $\cb^\nlo$ should be
determined free from the infrared divergence, 
we have to deal with it in the process of calculations.   
Following the method employed in Ref.~\cite{kura},
we subtract, from the original lattice integrand of each one-loop diagram,
a continuum-like integrand which is an analytically integrable expression 
with the same infrared behavior of the lattice integrand.
Suppose $\intlat I(k,\{p\},\lambda)$ is a lattice Feynman integral 
for a given one-loop diagram, whose infrared divergence is
regularized by  the fictitious gluon mass $\lambda$.
$\{p\}$ denotes a set of external momenta.
The infrared divergent term is extracted as
\ben
\intlat I(k,\{p\},\lambda)&=&\left.\intlat \left[I(k,\{p\},\lambda)
-{\tilde I}(k,\{p\},\lambda)\right]\right|_{\lambda\rightarrow 0}\nn\\
&&+\intlat {\tilde I}(k,\{p\},\lambda),
\een
where ${\tilde I}(k,\{p\},\lambda)$, which 
has the same infrared behavior with
$I(k,\{p\},\lambda)$, should be analytically integrable.
The infrared divergence is transferred in the last term.
A candidate of the counter integrand  
${\tilde I}(k,\{p\},\lambda)$ depends on each
one-loop diagram. We will explain it in the following sections.

\section{Determination of $\nu$ and $r_s$ at the one-loop level}

At the tree level the parameters are adjusted such that the 
quark propagator of eq.(\ref{eq:qprop}) reproduces 
the correct relativistic form\cite{akt}:
\be
S_q(p)=
\frac{1}{Z_q^\lo}
\frac{ -i \gamma_0 p_0- i \sum_i \gamma_i p_i + m_p^\lo}
{p_0^2+\sum_i p_i^2 +{m_p^\lo}^2}
+{\rm (no\;\; pole\;\; terms)}+O((p_i a)^2)
\label{eq:qprop_rel}
\ee
around the pole. $Z_q^\lo$ and $m_p^\lo$ are extracted with $p_i=0$,
\ben
m_p^\lo&=&{\rm log}\left|\frac{m_0+r_t+\sqrt{m_0^2+2r_t m_0+1}}
{1+r_t}\right|,
\label{eq:polemass_0}\\
Z_m^\lo&=&\frac{m_p^\lo}{m_0},\\
Z_q^\lo&=&{\rm cosh}(m_p^\lo)+r_t {\rm sinh}(m_p^\lo).
\een
Imposing finite spatial momenta 
we determine $\nu^\lo$ from the speed of light
and $r_s^\lo$ from the dispersion relation.
The results are given by
\ben
\nu^\lo&=&\frac{{\rm sinh}(m_p^\lo)}{m_p^\lo},
\label{eq:nu_qp}\\
r_s^\lo&=&\frac{{\rm cosh}(m_p^\lo)+r_t {\rm sinh}(m_p^\lo)}{m_p^\lo}
-\frac{{\rm sinh}(m_p^\lo)}{{m_p^\lo}^2}\\
&=&\frac{1}{m_p^\lo}(Z_q^\lo-\nu^\lo).
\label{eq:rs_qp}
\een

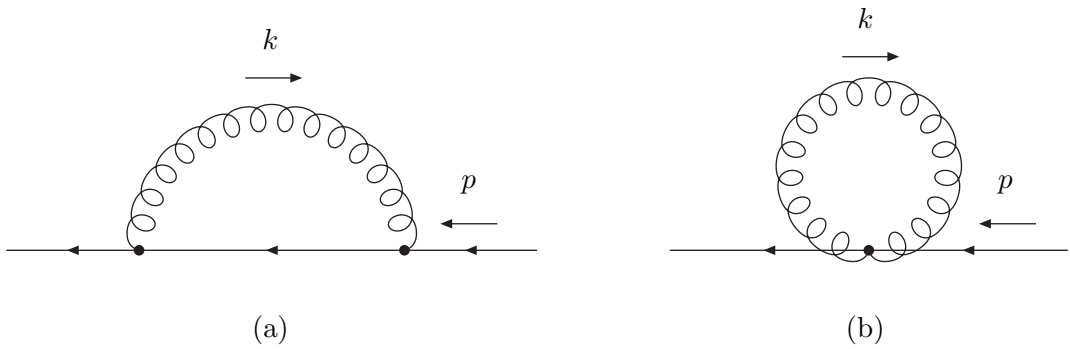
\begin{figure}[t]
\begin{center}

\begin{picture}(500,150)(0,0)
\ArrowLine(225,50)(175,50)
\Vertex(175,50){2}
\LongArrow(210,60)(190,60)
\Text(200,75)[c]{$p$}
\ArrowLine(175,50)(75,50)
\Vertex(75,50){2}
\ArrowLine(75,50)(25,50)
\GlueArc(125,50)(50,0,180){5}{14}
\LongArrow(115,115)(135,115)
\Text(125,130)[c]{$k$}
\Text(125,20)[c]{(a)}
\ArrowLine(425,50)(350,50)
\Vertex(350,50){2}
\LongArrow(413,60)(393,60)
\Text(403,75)[c]{$p$}
\ArrowLine(350,50)(275,50)
\GlueArc(350,80)(30,-90,270){5}{16}
\LongArrow(340,123)(360,123)
\Text(350,138)[c]{$k$}
\Text(350,20)[c]{(b)}
\end{picture}

\vspace{-4mm}
\end{center}
\caption{One-loop diagrams for the quark self-energy.}
\label{fig:qse_1loop}
\end{figure}

The one-loop contributions to the quark self-energy are depicted in
Fig.~\ref{fig:qse_1loop}, whose expression is given by
\ben
g^2\Sigma(p,m_0)&=&g^2\intlat I_\Sigma(k,p,m_0) \nn\\
&=&g^2
\left[i\gamma_0 \sin p_0 B_0(p,m_0)
+\nu i \sum_i\gamma_i \sin p_i B_i(p,m_0)+C(p,m_0)\right].
\een
Incorporating this contribution, the inverse quark propagator
up to the one-loop level is written as
\ben
S^{-1}(p,m)&=&i\gamma_0\sin p_0[1-g^2 B_0(p,m)]
+\nu i \sum_i\gamma_i\sin p_i[1-g^2 B_i(p,m)]+m\nn\\
&&+2 r_t\sin^2 \left(\frac{p_0}{2}\right)
+2 r_s\sum_i \sin^2 \left(\frac{p_i}{2}\right) 
-g^2{\hat C}(p,m),
\label{eq:qp_inv_1loop}
\een
where we redefine the quark mass as
\ben
m&=&m_0-g^2 C(p=0,m=0),\\
{\hat C}(p,m)&=&C(p,m)-C(p=0,m=0).
\een
With this definition 
the inverse quark propagator satisfies the on-shell condition  
for the massless quark up to the one-loop level : $S^{-1}(p_0=0,p_i=0,m=0)=0$. 
For convenience we replace 
the tree level pole mass of eq.(\ref{eq:polemass_0}) by
\ben
m_p^\lo&=&{\rm log}\left|\frac{m+r_t+\sqrt{m^2+2r_t m+1}}
{1+r_t}\right|,
\label{eq:polemass}
\een
where $m_p^\lo=0$ at $m=0$.
In the following analyses we use $m$ as if it were the ``bare'' quark mass. 

\begin{figure}[t]
\centering{
\hskip -0.0cm
\includegraphics[width=140mm,angle=0]{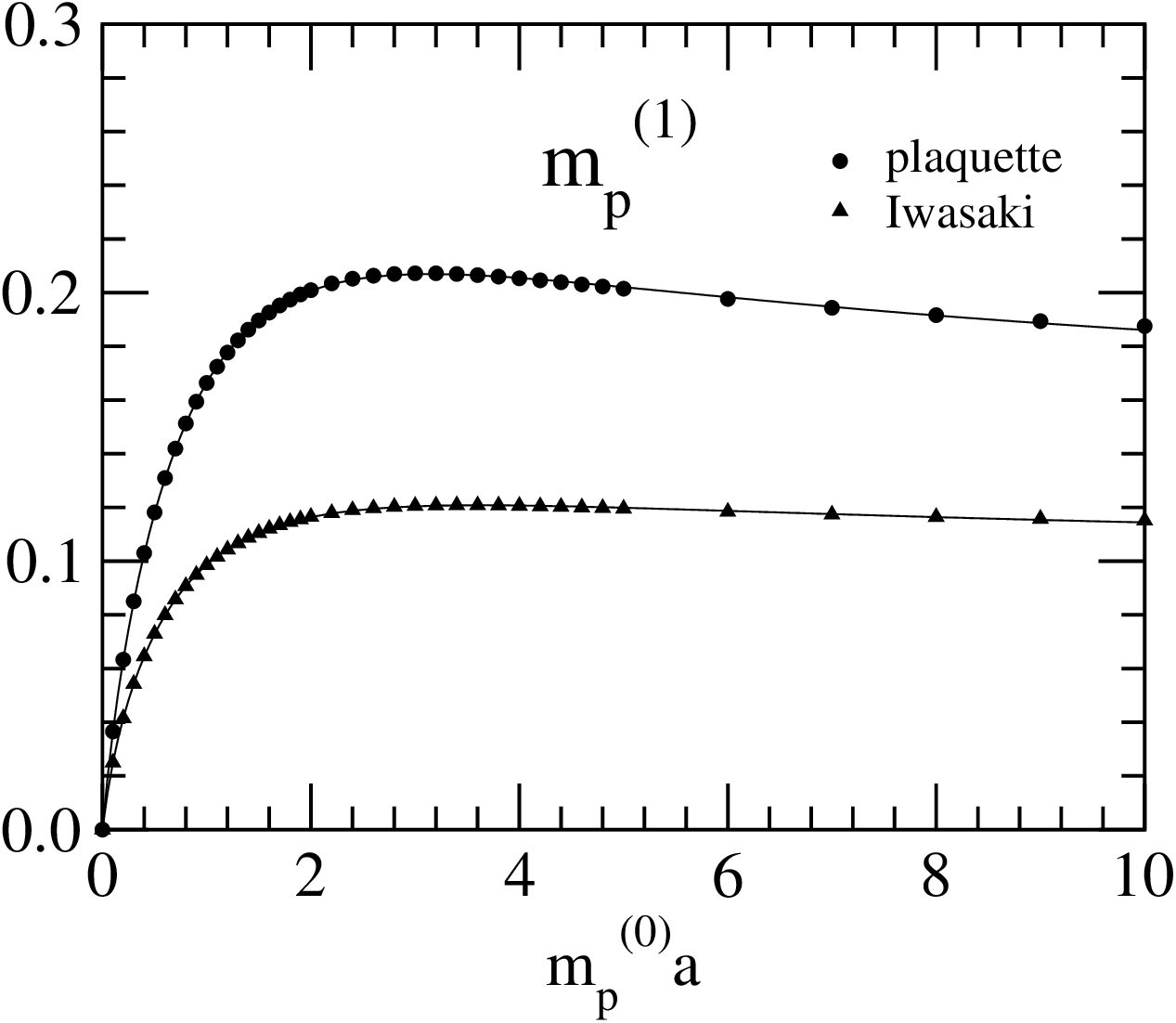}     
}
\caption{$m_p^\nlo$ as a function of $m_p^\lo$. Solid lines denote 
the interpolation of $m_p^\nlo$ with eq.(\protect{\ref{eq:fit_mp}})}
\label{fig:mp}
\vspace{8mm}
\end{figure}

The pole mass $m_p$ is obtained from the pole of the quark propagator: 
$S^{-1}(p_0=i m_p, p_i=0 ,m)=0$.
We obtain
\ben
m_p^\nlo&=&-\frac{1}{Z_q^\lo}{\rm Tr}
\left[\frac{(\gamma_0+1)}{4}
\left\{\Sigma(\pos,m)-\Sigma(p=0,m=0)\right\}\right],
\een
where $\pos=(p_0=im_p^\lo,p_i=0)$. It is noted that 
$m_p^\nlo$ has no infrared divergence.
In Fig.~\ref{fig:mp} 
we plot the $m_p^\lo$ dependence of $m_p^\nlo$ for the plaquette and the 
Iwasaki gauge actions. We observe that $m_p^\nlo$ vanishes
at $m_p^\lo$=0 as expected.  The solid lines denote the fitting results
of the parameterization:
\ben
m_p^\nlo=
\frac{\sum_{i=1}^3 a_i\{\mplo\}^i}{1+\sum_{i=1}^3 b_i\{\mplo\}^i}.
\label{eq:fit_mp}
\een
The relative errors of this interpolation are less than 1\% 
over the range $0< \mplo\le 10$.
We tabulate the values of the parameters $a_i$ and $b_i$ ($i=1,\dots,3$)
in Table~\ref{tab:fit_mp}.

The wave function $Z_q$ is defined 
as the residue of the quark propagator $S(p,m)$.
The one-loop contribution on the lattice is given by
\ben
Z_q^\nlo=\{\sinh(m_p^\lo)+r_t\cosh(m_p^\lo)\}m_p^{\nlo}
-{\rm Tr}\left[\frac{(\gamma_0+1)}{4}(-i)
\frac{\partial\Sigma_\latt}{\partial p_0}(\pos,m)\right].
\label{eq:wf_latt_1l}
\een
The infrared divergence in $Z_q^\nlo$ 
is regularized by introducing the fictitious gluon mass $\lambda$. 
We extract the divergent term in $Z_q^\nlo$ by subtracting from 
$I_{\Sigma}$ an analytically integrable expression 
${\tilde I}_{\Sigma}$ which has the same infrared behavior as $I_\Sigma$.
As a candidate of ${\tilde I}_{\Sigma}$ we take
\ben
{\tilde I}_{\Sigma}(k,p,m_p^\lo)=C_F Z_q^\lo \theta(\Lambda^2-k^2)
i\gamma_\alpha \frac{1}{i(\pslash+\kslash)+m_p^\lo}
i\gamma_\alpha \frac{1}{k^2+\lambda^2}
\label{eq:ti_sigma}
\een
with  a cut-off $\Lambda$ $(\le\pi)$.
The integration is easily performed\cite{kura}
\ben
&&\intlat \frac{\partial {\tilde I}_\Sigma}
{i\partial \pslash}(k,\pos,\mplo)
\nn \\
&=&Z_q^\lo \frac{C_F}{16\pi^2}\left[
-2\log\left|{\frac{\lambda^2}{\Lambda^2}}\right|
-\frac{3}{4}\frac{\Lambda^4}{\mplo^4}-\frac{9}{2}\frac{\Lambda}
{\mplo^2}\sqrt{\Lambda^2+4\mplo^2}
\right.\nn \\ 
&&+\frac{3}{4}\frac{\Lambda}{\mplo^4}(\Lambda^2+4\mplo^2)^{\frac{3}{2}}
\left.-6\log\left|{\frac{\Lambda+\sqrt{\Lambda^2+4\mplo^2}}
{2\mplo}}\right|\right],
\label{eq:int_wf_latt}
\een
It is noted that the divergent term 
is the same as the Wilson case in Ref.~\cite{kura}.

\begin{figure}[b]
\centering{
\hskip -0.0cm
\includegraphics[width=140mm,angle=0]{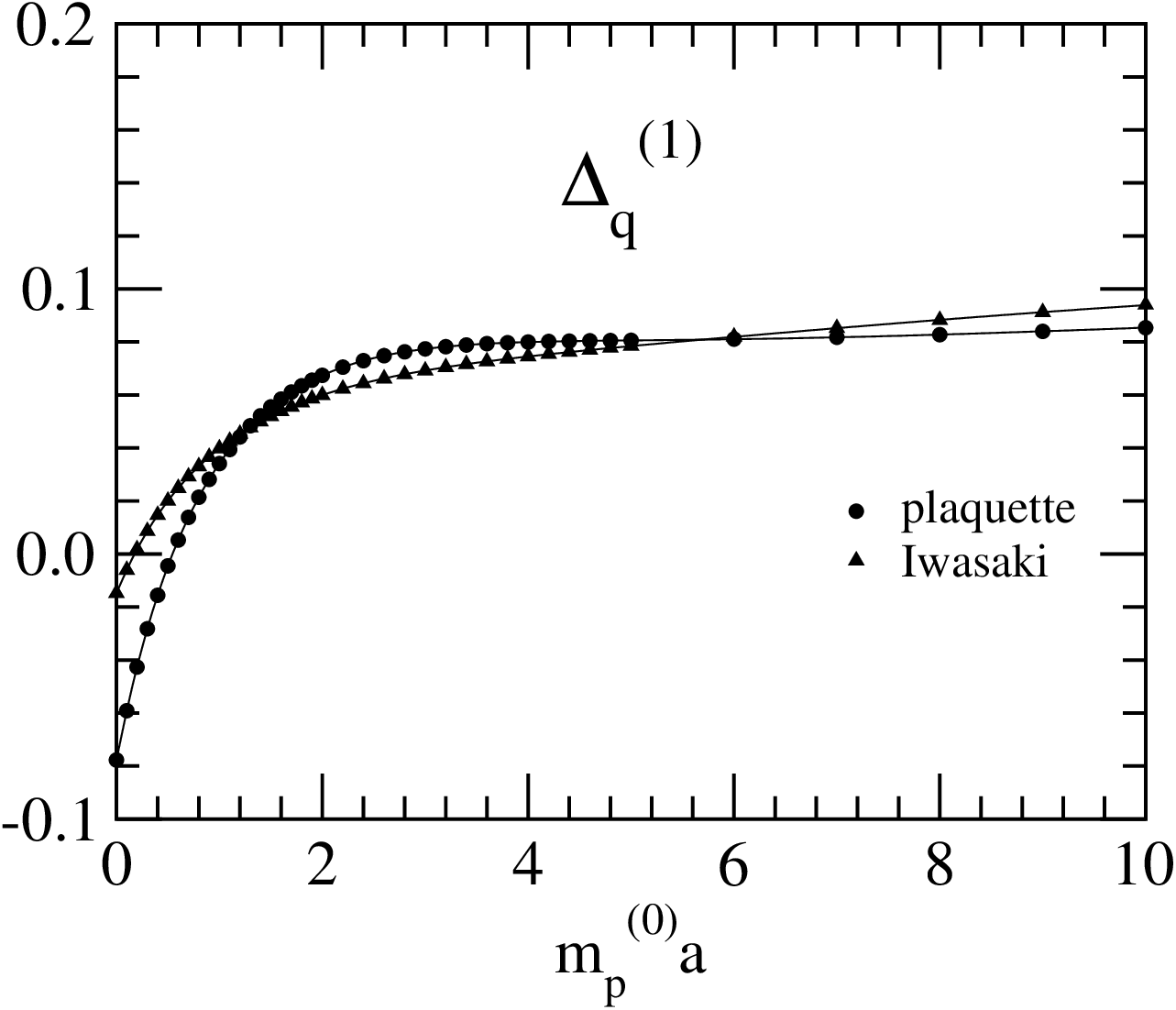}     
}
\caption{$\Delta_q^\nlo$ as a function of $m_p^\lo$. Solid lines denote 
the interpolation of $\Delta_q^\nlo$ with eq.(\protect{\ref{eq:fit_wf}})}
\label{fig:wf}
\vspace{8mm}
\end{figure}

The finite renormalization factor from the lattice regularization scheme
to the continuum Naive Dimensional Regularization (NDR) scheme 
($\psi_{\cont} = Z_q^{-\frac{1}{2}} \psi_{\latt}$)
is determined by 
\ben
Z_q &\equiv& \left(Z_{q,\latt}^\lo\right)^{-1} 
\left[1-g^2\Delta_q^\nlo(\mplo)\right] = \frac{Z_{q,\cont}}{Z_{q,\latt}}
\een 
up to the one-loop level. 
Here the continuum wave function renormalization factor is given by
\ben
Z_{q,\cont}=1-g^2 
\frac{\partial\Sigma_\cont}{i\partial \pslash}(\pos,\mplo),
\label{eq:wf_cont_1l}
\een
with
\ben
\Sigma_\cont(p,\mplo)
=C_F\int_{-\infty}^{\infty}\frac{d^D k}{(2\pi)^D}
i\gamma_\alpha \frac{1}{i(\pslash+\kslash)+\mplo}
i\gamma_\alpha \frac{1}{k^2+\lambda^2},
\een
where $D=4-2\epsilon$ ($\epsilon>0$) in the NDR scheme. 
The momentum assignment is depicted in Fig.~\ref{fig:qse_1loop} (a).
After some algebra we obtain
\ben
\frac{\partial\Sigma_\cont}{i\partial \pslash}(\pos,\mplo)
=\frac{C_F}{16\pi^2}\left[-\frac{1}{\bar\epsilon}
-\log\left|\frac{\mu^2}{{\mplo}^2}\right|
-2\log\left|\frac{\lambda^2}{{\mplo}^2}\right|-4\right],
\label{eq:int_wf_cont}
\een
where $1/\bar\epsilon=1/\epsilon-\gamma+\ln(4\pi)$ and $\mu$ is
the renormalization scale.
In the $\msbar$ scheme, the pole term $1/\bar\epsilon$ 
should be eliminated.
 From eqs.(\ref{eq:wf_latt_1l}) and (\ref{eq:wf_cont_1l}) 
the finite renormalization constant
$\Delta_q^\nlo(\mplo)$ is expressed as
\ben
\Delta_q^\nlo(\mplo)&=&
\frac{\partial\Sigma_\cont}{i\partial \pslash}(\pos,\mplo)\nn\\
&&\left.+\frac{\sinh(\mplo)+r_t\cosh(\mplo)}{Z_q^\lo}m_p^{\nlo}
-\frac{1}{Z_q^\lo}{\rm Tr}\left[\frac{(\gamma_0+1)}{4}(-i)
\frac{\partial\Sigma_\latt}{\partial p_0}(\pos,m)\right]\right|_{\lambda=0}.
\een
Comparing eqs.(\ref{eq:int_wf_latt}) and (\ref{eq:int_wf_cont})
we find that the infrared divergence for $\lambda\rightarrow 0$
and the mass singularities
at $\mplo\rightarrow 0$ are exactly canceled out, which assures that
$\Delta_q^\nlo(\mplo)$ is finite even in the massless limit.
Figure~\ref{fig:wf} shows the $\mplo$ dependence of 
$\Delta_q^\nlo(\mplo)$ for the plaquette and the 
Iwasaki gauge actions. We parameterize $\Delta_q^\nlo$ as
\ben
\Delta_q^\nlo=\Delta_q^\nlo(\mplo=0)+
\frac{\sum_{i=1}^4 a_i\{\mplo\}^i}{1+\sum_{i=1}^4 b_i\{\mplo\}^i},
\label{eq:fit_wf}
\een
where the values of $\Delta_q^\nlo(\mplo=0)$ 
except for the DBW2 gauge action are taken from Ref.~\cite{gmass}. 
The fitting results are drawn in Fig.~\ref{fig:wf} by solid lines, whose
relative errors are at most 1\%  over the range $0< \mplo\le 10$.
The values of the parameters $a_i$,  $b_i$ ($i=1,\dots,4$) 
and $\Delta_q^\nlo(\mplo=0)$ are listed in Table~\ref{tab:fit_wf}.

\begin{figure}[t]
\centering{
\hskip -0.0cm
\includegraphics[width=140mm,angle=0]{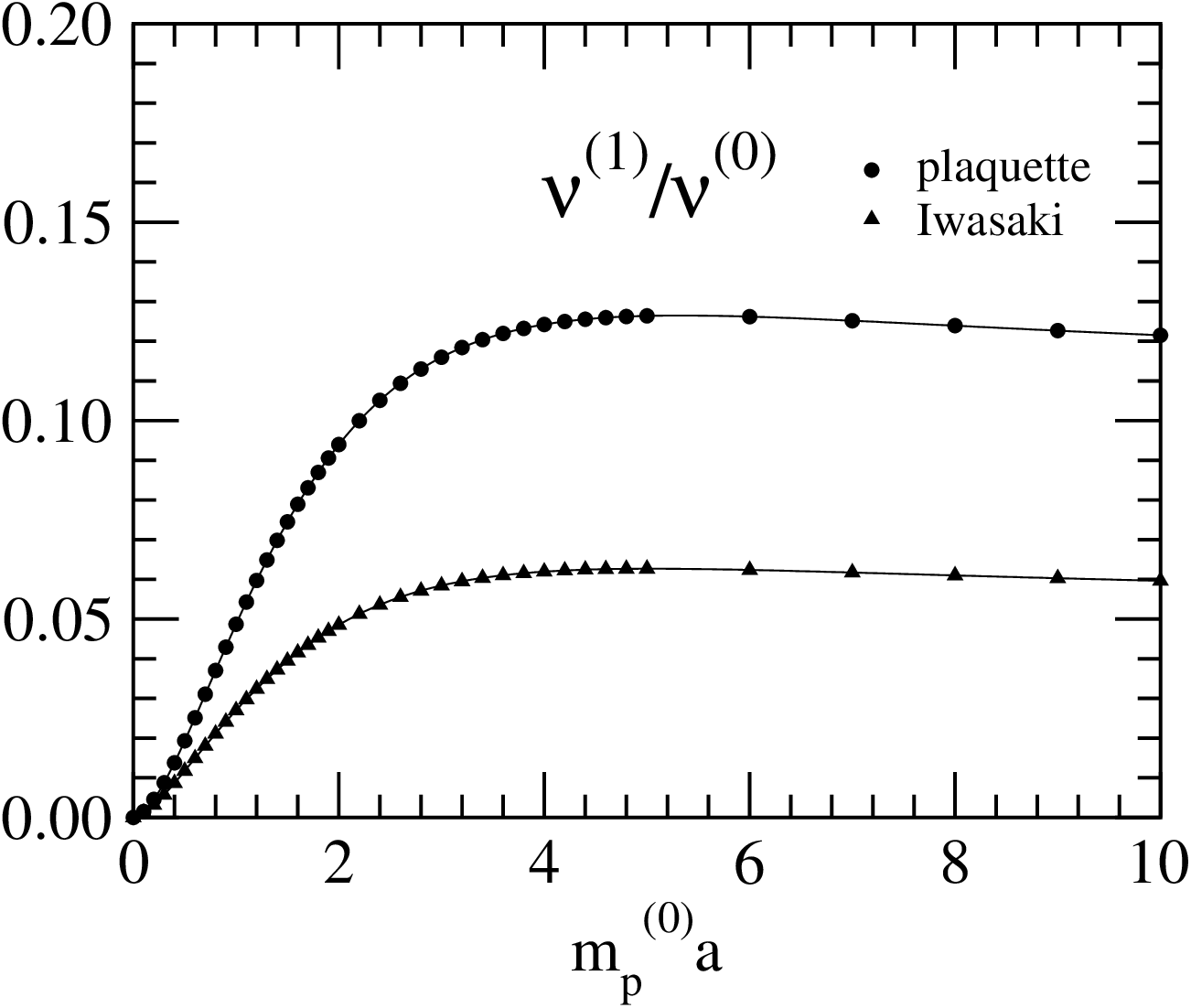}     
}
\caption{$\nu^\nlo/\nu^\lo$ as a function of $m_p^\lo$. Solid lines denote 
the interpolation of $\nu^\nlo/\nu^\lo$ with eq.(\protect{\ref{eq:fit_nu}})}
\label{fig:nu}
\vspace{8mm}
\end{figure}

The parameter $\nu$ is determined by adjusting the speed of light
in $S^{-1}(p,m)$. Comparing the coefficients of $\gamma_0$ 
and $\gamma_i$ in the numerator we obtain
\ben
\nu=\frac{\sinh(m_p)}{m_p}
\frac{[1-g^2 B_0(\pos,m)]}{[1-g^2 B_i(\pos,m)]}.
\een
The one-loop contribution is given by
\ben
\nu^\nlo=\left( \frac{\cosh(\mplo)}{\mplo}
-\frac{\sinh(\mplo)}{(\mplo)^2}\right)m_p^\nlo
+\nu^\lo\{B_i(\pos,m)-B_0(\pos,m)\},
\een
where $B_i$ and $B_0$ have no infrared divergence.
The quark mass dependences of $\nu^\nlo/\nu^\lo$ for the plaquette and the 
Iwasaki gauge actions are shown in Fig.~\ref{fig:nu}.
As expected $\nu^\nlo$ vanishes at $m_p^\lo=0$ for both cases.
The solid lines depict the results of the interpolation:
\ben
\frac{\nu^\nlo}{\nu^\lo}=
\frac{\sum_{i=1}^5 a_i\{\mplo\}^i}{1+\sum_{i=1}^5 b_i\{\mplo\}^i}.
\label{eq:fit_nu}
\een
The relative errors of this interpolation are less than a few \% 
over the range $0< \mplo\le 10$.
The values of the parameters $a_i$ and $b_i$ ($i=1,\dots,5$)
are collected in Table~\ref{tab:fit_nu}.

\begin{figure}[b]
\centering{
\hskip -0.0cm
\includegraphics[width=140mm,angle=0]{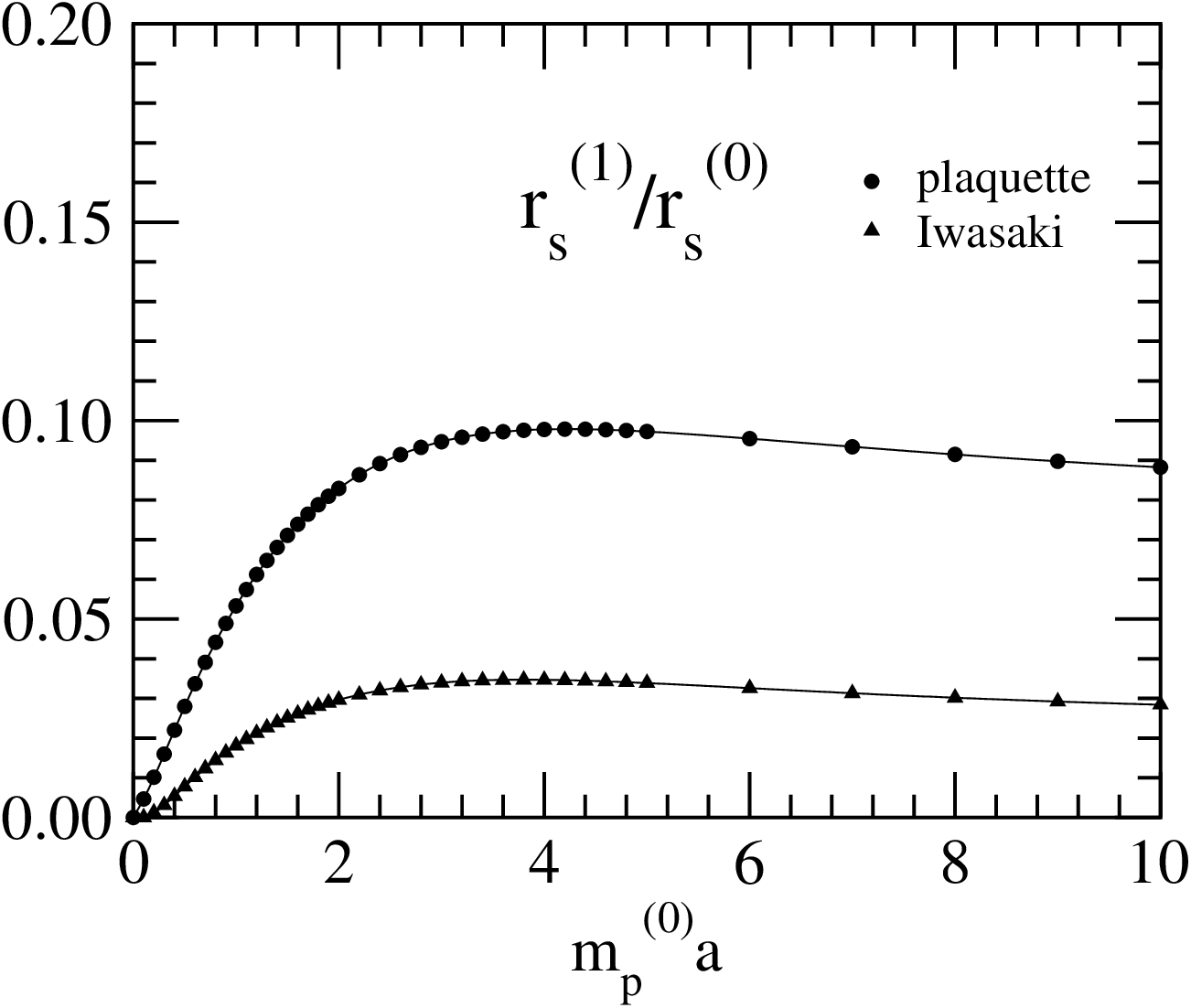}     
}
\caption{$r_s^\nlo/r_s^\lo$ as a function of $m_p^\lo$. Solid lines denote 
the interpolation of $r_s^\nlo/r_s^\lo$ with eq.(\protect{\ref{eq:fit_rs}})}
\label{fig:rs}
\vspace{8mm}
\end{figure}

The parameter $r_s$ is determined from $S^{-1}(p,m)$ 
such that the correct dispersion relation is reproduced:
\ben
E^2=m_p^2+\sum_i p_i^2+O(p_i^4).
\een
This condition yields
\ben
r_s^\nlo&=&\frac{1}{m_p^\lo}\{Z_q^\nlo
+\nu^\lo B_i(\pos,m)-\nu^\nlo\}
-\frac{m_p^\nlo}{m_p^\lo}r_s^\lo
+{\rm Tr}\left[\frac{(1+\gamma_0)}{2}
\frac{\partial\Sigma}{\partial p_k^2}(\pos,m)\right].
\een
The infrared divergence of the last term can be extracted 
by using ${\tilde I}_{\Sigma}$ in eq.(\ref{eq:ti_sigma}):
\ben
&&\intlat \frac{\partial {\tilde I}_\Sigma}
{\partial p_k^2}(k,\pos,\mplo)
\nn \\
&=&\frac{Z_q^\lo}{\mplo}\frac{C_F}{16\pi^2}\left[
\log\left|\frac{\Lambda^2}{\lambda^2}\right|-\frac{\Lambda^2}{\mplo^2}
-\frac{1}{2}\frac{\Lambda^4}{\mplo^4}-2\frac{\Lambda}
{\mplo^2}\sqrt{\Lambda^2+4\mplo^2}
\right.\nn \\ 
&&+\frac{1}{2}\frac{\Lambda}{\mplo^4}(\Lambda^2+4\mplo^2)^{\frac{3}{2}}
\left.-2\log\left|{\frac{\Lambda+\sqrt{\Lambda^2+4\mplo^2}}
{2\mplo}}\right|\right].
\label{eq:int_rs_latt}
\een
We find that the infrared divergence and the mass singularity 
in the last term are exactly canceled out by those in $Z_q^\nlo/{\mplo}$.
We show the $\mplo $ dependence of $r_s^\nlo/r_s^\lo$ for the plaquette
and the Iwasaki gauge actions
in Fig.~\ref{fig:rs}, where $r_s^\nlo/r_s^\lo$ 
becomes close to zero as $\mplo$ vanishes. 
This is an expected
behavior because the deviation of $r_s$ from $r_t$ stems from
the power corrections of $\mplo a$. 
The quark mass dependence of $r_s^\nlo/r_s^\lo$ 
is well described by the interpolation with the relative errors of 
less than a few \% over the range $0< \mplo\le 10$, 
\ben
\frac{r_s^\nlo}{r_s^\lo}=
\frac{\sum_{i=1}^5 a_i\{\mplo\}^i}{1+\sum_{i=1}^5 b_i\{\mplo\}^i}.
\label{eq:fit_rs}
\een
We give the values of the parameters $a_i$ and $b_i$ ($i=1,\dots,5$)
in Table~\ref{tab:fit_rs}.

\section{Determination of $\ce$ and $\cb$ up to the one-loop level}

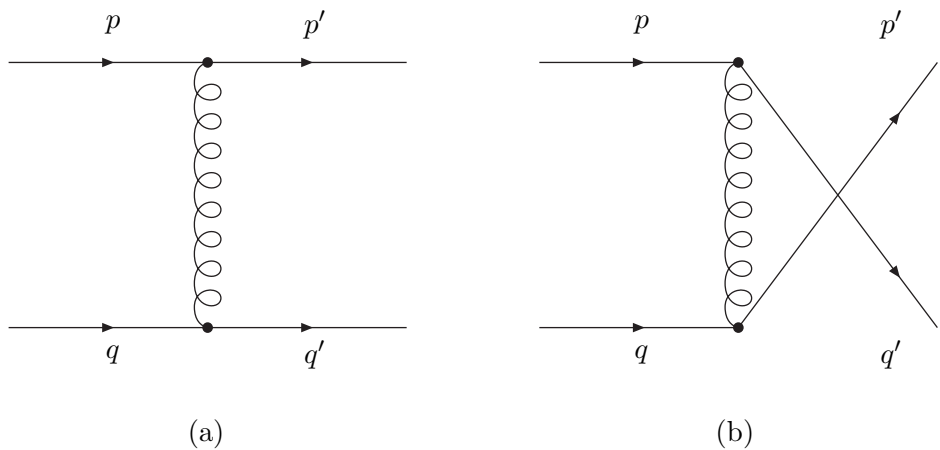
\begin{figure}[t]
\begin{center}

\begin{picture}(450,180)(0,0)
\ArrowLine(50,150)(125,150)
\Text(87,165)[l]{$p$}
\Vertex(125,150){2}
\ArrowLine(125,150)(200,150)
\Text(162,165)[l]{$p^\prime$}
\ArrowLine(50,50)(125,50)
\Text(87,40)[l]{$q$}
\Vertex(125,50){2}
\ArrowLine(125,50)(200,50)
\Text(162,40)[l]{$q^\prime$}
\Gluon(125,150)(125,50){-5}{8}
\Text(125,10)[c]{(a)}
\ArrowLine(250,150)(325,150)
\Text(287,165)[l]{$p$}
\Vertex(325,150){2}
\LongArrow(325,150)(385,70)
\Line(385,70)(400,50)
\Text(380,165)[l]{$p^\prime$}
\ArrowLine(250,50)(325,50)
\Text(287,40)[l]{$q$}
\Vertex(325,50){2}
\LongArrow(325,50)(385,130)
\Line(385,130)(400,150)
\Text(380,40)[l]{$q^\prime$}
\Gluon(325,150)(325,50){-5}{8}
\Text(325,10)[c]{(b)}
\end{picture}

\end{center}
\caption{Tree level diagrams for the quark-quark scattering.}
\label{fig:scatt_tree}
\vspace{8mm}
\end{figure}

We employ the on-shell quark-quark scattering
amplitude to determine $\ce$ and $\cb$.
At the tree level the parameters $\nu$, $r_s$, $\ce$ and $\cb$
are adjusted to reproduce the continuum form of
the scattering amplitude at the on-shell point 
removing the $m_Q a$ corrections\cite{akt},
\ben
T &=&-g^2(T^A)^2{\bar u}(p^\prime)\gamma_\mu u(p)D_{\mu\nu}(p-p^\prime)
{\bar u}(q^\prime)\gamma_\nu u(q)\nn\\
&&-g^2(T^A)^2{\bar u}(q^\prime)\gamma_\mu u(p)D_{\mu\nu}(p-q^\prime)
{\bar u}(p^\prime)\gamma_\nu u(q)\nn\\
&&+O((p_i a)^2,(q_i a)^2,(p^\prime_i a)^2,(q_i^\prime a)^2),
\een
where the momentum assignment is depicted in Fig.~\ref{fig:scatt_tree}
and $D_{\mu\nu}$ denotes the gluon propagator.
At the tree level the quark-quark-gluon vertex is written as
\ben
\left({\bar u}(p^\prime)\Lambda_0^\lo(p,p^\prime) u(p)\right)_\latt
&=&Z_q^\lo\left({\bar u}(p^\prime)i\gamma_0 u(p)\right)_\cont
+O((p_i a)^2,(p_i^\prime a)^2),
\label{eq:vtx_nrm_tree_t}\\
\left({\bar u}(p^\prime)\Lambda_k^\lo(p,p^\prime) u(p)\right)_\latt
&=&Z_q^\lo\left({\bar u}(p^\prime)i\gamma_k u(p)\right)_\cont
+O((p_i a)^2,(p_i^\prime a)^2),
\label{eq:vtx_nrm_tree_s}
\een
for 
\ben
\Lambda_0^\lo(p,p^\prime)
&=&i\gamma_0\cos\left(\frac{p_0+p^\prime_0}{2}\right)\
+r_t \sin\left(\frac{p_0+p^\prime_0}{2}\right)\nn\\
&&+\frac{\ce^\lo}{2} \cos\left(\frac{p_0-p^\prime_0}{2}\right)
\sum_l \sigma_{0l}\sin (p_l-p^\prime_l),\\
\Lambda_k^\lo(p,p^\prime)
&=&i\nu^\lo \gamma_k\cos \left(\frac{p_k+p^\prime_k}{2}\right)\
+r_s^\lo \sin \left(\frac{p_k+p^\prime_k}{2}\right)\nn\\
&&+\frac{\ce^\lo}{2} \cos\left(\frac{p_k-p^\prime_k}{2}\right)
\sigma_{k0}\sin (p_0-p^\prime_0)\nn\\
&&+\frac{\cb^\lo}{2} \cos\left(\frac{p_k-p^\prime_k}{2}\right)
\sum_{l\ne k} \sigma_{kl}\sin(p_l-p^\prime_l),
\een
where the spinor on the lattice is given by
\be
u(p)=\left( \begin{array}{c}\phi\\
\frac{\nu{\vec p}\cdot{\vec \sigma}}{N(p)}\phi
\end{array}\right)
+O((p_i a)^2),
\ee
with $N(p)=(-i){\rm sin}(p_0)+m_0+r_t(1-{\rm cos}(p_0))
+r_s\sum_i(1-{\rm cos}(p_i))$.
The $O(a)$ improvement condition yields
\ben
\nu^\lo&=&\frac{\sinh(\mplo)}{\mplo},
\label{eq:nu_me}\\
r_s^\lo&=&\frac{\cosh(\mplo)+r_t \sinh(\mplo)}{\mplo}
-\frac{\sinh(\mplo)}{\mplo^2},
\label{eq:rs_me}\\
\ce^\lo&=&r_t \nu^\lo,
\label{eq:ce}\\
\cb^\lo&=&r_s^\lo.
\label{eq:cb}
\een
It should be noted that the values of 
$\nu^\lo$ and $r_s^\lo$ are exactly the 
same as those determined from the quark propagator.

\begin{figure}[t]
\begin{center}

\begin{picture}(450,150)(0,0)
\ArrowLine(50,50)(75,75)
\Text(70,55)[l]{$p$}
\ArrowLine(75,75)(100,100)
\Text(90,70)[l]{$p+k$}
\Vertex(75,75){2}
\ArrowLine(100,100)(75,125)
\Text(70,145)[l]{$q$}
\ArrowLine(75,125)(50,150)
\Text(90,130)[l]{$q+k$}
\Vertex(75,125){2}
\Vertex(100,100){2}
\Gluon(100,100)(200,100){5}{8}
\Text(150,130)[c]{$p-q$, $\mu$, $A$}
\LongArrow(140,115)(160,115)
\Vertex(75,75){2}
\Vertex(75,125){2}
\Gluon(75,75)(75,125){5}{4}
\LongArrow(60,110)(60,90)
\Text(50,100)[c]{$k$}
\Text(125,30)[c]{(a)}
\ArrowLine(250,50)(275,75)
\Gluon(300,100)(275,75){5}{3}
\Text(270,55)[l]{$p$}
\Text(290,70)[l]{$p-q+k$}
\Vertex(275,75){2}
\Gluon(275,125)(300,100){5}{3}
\ArrowLine(275,125)(250,150)
\Text(270,145)[l]{$q$}
\Text(290,130)[l]{$k$}
\Vertex(275,125){2}
\Vertex(300,100){2}
\Gluon(300,100)(400,100){5}{8}
\Text(350,130)[c]{$p-q$, $\mu$, $A$}
\LongArrow(340,115)(360,115)
\Vertex(275,75){2}
\Vertex(275,125){2}
\ArrowLine(275,75)(275,125)
\Text(270,100)[r]{$q-k$}
\Text(325,30)[c]{(b)}
\end{picture}

\begin{picture}(450,150)(0,0)
\ArrowLine(50,50)(100,100)
\Text(80,65)[l]{$p$}
\ArrowLine(100,100)(50,150)
\Text(80,135)[l]{$q$}
\Vertex(100,100){2}
\LongArrow(130,130)(110,130)
\Text(120,140)[c]{$k$}
\GlueArc(120,100)(20,0,180){5}{7}
\GlueArc(120,100)(20,180,360){5}{7}
\LongArrow(110,70)(130,70)
\Text(120,60)[c]{$p-q+k$}
\Vertex(140,100){2}
\Gluon(140,100)(200,100){5}{5}
\LongArrow(160,115)(180,115)
\Text(170,130)[c]{$p-q$, $\mu$, $A$}
\Text(125,30)[c]{(c)}
\ArrowLine(250,50)(300,100)
\Text(280,65)[l]{$p$}
\ArrowLine(300,100)(250,150)
\Text(280,135)[l]{$q$}
\Vertex(300,100){2}
\LongArrow(330,130)(310,130)
\Text(320,140)[c]{$k$}
\GlueArc(320,100)(20,-180,180){5}{14}
\GlueArc(350,100)(50,180,360){5}{16}
\LongArrow(380,115)(400,115)
\Text(390,130)[c]{$p-q$, $\mu$, $A$}
\Text(325,30)[c]{(d)}
\end{picture}

\begin{picture}(450,150)(0,0)
\ArrowLine(50,50)(75,75)
\Vertex(75,75){2}
\Text(45,100)[r]{$q+k$}
\ArrowArcn(75,100)(25,270,90)
\Text(70,55)[l]{$p$}
\ArrowLine(75,125)(50,150)
\Text(70,145)[l]{$q$}
\Vertex(75,125){2}
\GlueArc(75,100)(25,-90,90){5}{8}
\LongArrow(110,110)(110,90)
\Text(115,100)[l]{$k$}
\Gluon(200,75)(75,75){5}{11}
\Text(160,105)[c]{$p-q$, $\mu$, $A$}
\LongArrow(150,90)(170,90)
\Text(125,30)[c]{(e)}
\ArrowLine(250,50)(275,75)
\Vertex(275,75){2}
\ArrowArcn(275,100)(25,270,90)
\Text(270,55)[l]{$p$}
\ArrowLine(275,125)(250,150)
\Text(270,145)[l]{$q$}
\Vertex(275,125){2}
\GlueArc(275,100)(25,-90,90){5}{8}
\Text(245,100)[r]{$p+k$}
\LongArrow(310,110)(310,90)
\Text(315,100)[l]{$k$}
\Gluon(275,125)(400,125){5}{11}
\Text(360,95)[c]{$p-q$, $\mu$, $A$}
\LongArrow(350,110)(370,110)
\Text(325,30)[c]{(f)}
\end{picture}

\vspace{-8mm}
\end{center}
\caption{Quark-gluon vertex at the one-loop level.}
\label{fig:vtx_1loop}
\end{figure}
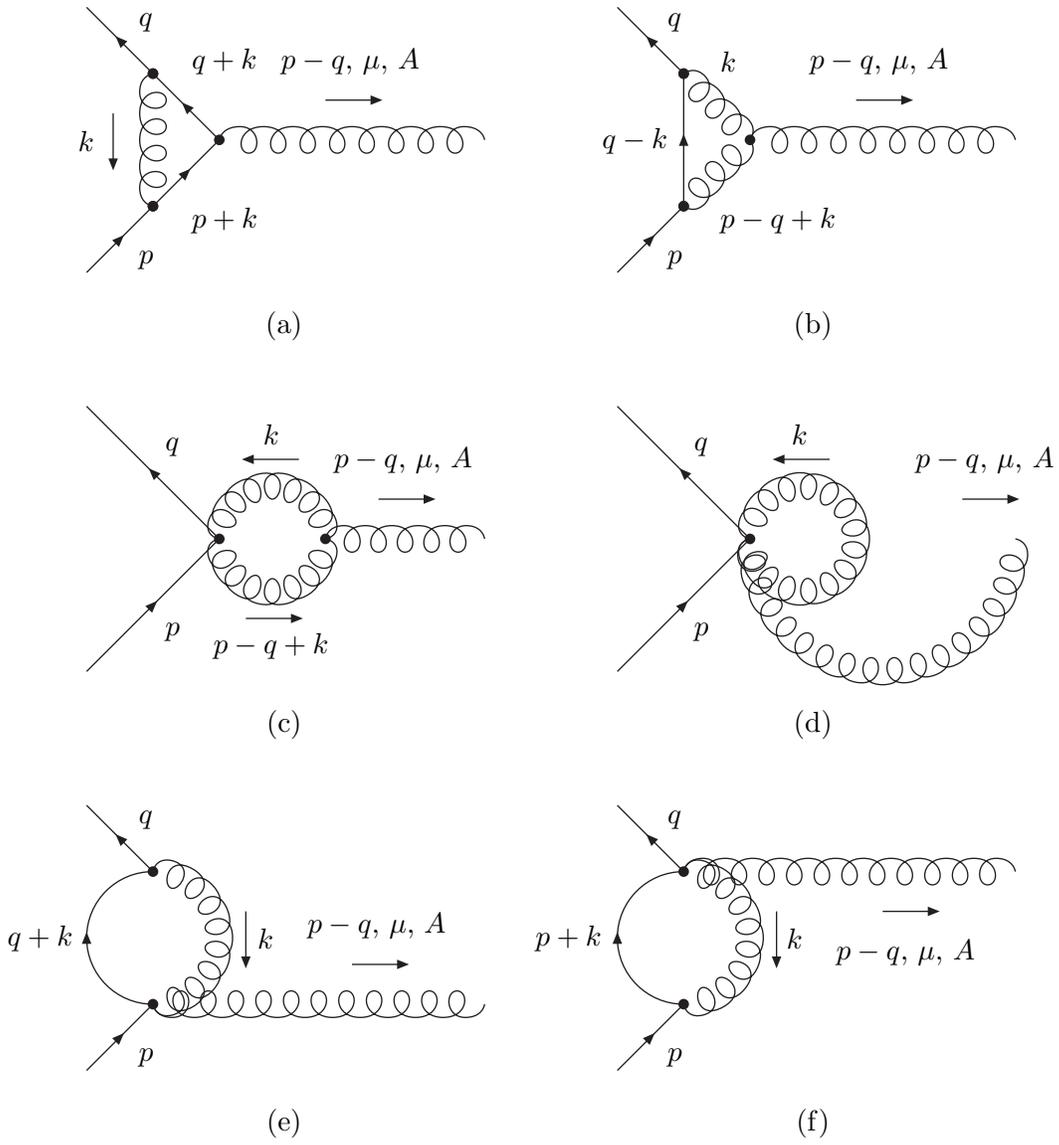

Let us turn to the one-loop calculation. 
Recently the authors have shown the validity of 
the conventional perturbative method 
to determine the clover coefficient $\csw$ up to the one-loop level
in the massless case from the on-shell quark-quark
scattering amplitude\cite{csw_m0}.
We extend this calculation to the massive case.
According to Ref.~\cite{csw_m0}, it is sufficient for us 
to improve each on-shell quark-quark-gluon vertex individually.
To determine the one-loop coefficients $\ce^\nlo$ and $\cb^\nlo$
we need six types of diagrams shown in Fig.~\ref{fig:vtx_1loop}.
We first consider to calculate $\cb^\nlo$.
Without the space-time symmetry the general form of the off-shell vertex
function at the one-loop level is written as
\ben
\Lambda_k^{(1)}(p,q,m) 
&=&\sum_{i=a,\dots,f}\Lambda_k^{(1-i)}(p,q,m)\nn\\
&=&\sum_{i=a,\dots,f}\int_{-\pi}^{\pi}\frac{d^4 k}{(2\pi)^4}
I_k^{(i)}(k,p,q,m)\nn\\
&=&\gamma_k F_1^k
+\gamma_k\{\pslash F_2^k+\pslash_0 F_3^k\}
+\{\qslash F_4^k+\qslash_0 F_5^k\}\gamma_k\nn\\
&&+\qslash\gamma_k\pslash F_6^k
+\qslash\gamma_k\pslash_0 F_7^k
+\qslash_0\gamma_k\pslash F_8^k
\nn\\
&&+(p_k+q_k)\left[ H_1^k+\pslash H_2^k
+\qslash H_3^k+\qslash\pslash H_4^k\right]\nn\\
&&+(p_k-q_k)\left[ G_1^k+\pslash G_2^k
+\qslash G_3^k+\qslash\pslash G_4^k\right]+O(a^2),
\label{eq:vtx_k_offsh}
\een  
where $\Lambda_k(p,q,m)=\Lambda_k^\lo(p,q,m)
+g^2\Lambda_k^\nlo(p,q,m)+O(g^4)$ and  
\ben
\pslash&=&\sum_{\alpha=0}^3 p_\alpha \gamma_\alpha, \\
\qslash&=&\sum_{\alpha=0}^3 q_\alpha \gamma_\alpha, \\
\pslash_0&=& p_0 \gamma_0, \\
\qslash_0&=& q_0 \gamma_0.
\een
The coefficients $F_i^k$ ($i=1,\dots,8$), $G_i^k$ ($i=1,2,3,4$) and 
$H_i^k$ ($i=1,2,3,4$) are functions of 
$p^2$, $q^2$, $p\cdot q$ and $m$.
 From the charge conjugation symmetry they have to satisfy
the following condition:
\ben
&&F_2^k=F_4^k,\\
&&F_3^k=F_5^k,\\
&&F_7^k=F_8^k,\\
&&H_2^k=H_3^k,\\
&&G_1^k=G_4^k=0,
\label{eq:g1kg4k}\\
&&G_2^k=-G_3^k.
\label{eq:g2kg3k}
\een
Sandwiching $\Lambda_k^{(1)}(p,q,m)$ by the on-shell quark states
$u(p)$ and $\bar u(q)$, which satisfy
$\pslash u(p) = i m_p u(p)$ and $\bar u(q) \qslash = im_p \bar u(q)$,
the matrix element is reduced to 
\ben
&&{\bar u}(q)\Lambda_k^{(1)} (p,q,m)u(p)\nn\\
&=&{\bar u}(q)\gamma_k u(p)
\left\{F_1^k+i m_p (F_2^k+F_4^k)-m_p^2 F_6^k\right\} \nn\\
&&+{\bar u}(q)\gamma_k\gamma_0 u(p)(p_0-q_0)
\left\{F_3^k+i m_p F_7^k\right\} \nn\\
&&+(p_k+q_k){\bar u}(q)u(p)
\left\{H_1^k+im_p (H_2^k+H_3^k)-m_p^2 H_4^k\right\}\nn \\
&&+(p_k-q_k){\bar u}(q)u(p)
\left\{G_1^k+im_p (G_2^k+G_3^k)-m_p^2 G_4^k\right\}
+O(a^2),
\label{eq:vtx_k_onsh}
\een
where we use $F_3^k=F_5^k$ and $F_7^k=F_8^k$.
(Note that we can replace $m_p$ with $\mplo$ in the 1-loop diagrams.)
The first term in the right hand side contributes 
to the renormalization factor of the quark-quark-gluon vertex,
which is equal to $Z_q^\lo$ at the tree level.
 From eqs.(\ref{eq:g1kg4k}) and (\ref{eq:g2kg3k}) we find that the
last term of eq.(\ref{eq:vtx_k_onsh}) vanishes: 
this term is not allowed from the charge conjugation symmetry.
It is also possible 
to numerically check $G_1^k+i m_p (G_2^k+G_3^k)-m_p^2 G_4^k=0$. 

The contribution of the second term is $O(a^2)$.
This can be shown as follows. For simplicity we first consider
the case of $\lqcd\ll m_Q\ll a^{-1}$.
The difference of $p_0$ and $q_0$ is expressed as
\ben
(p_0-q_0)a&=&\left(\sqrt{m_p^2+p_i^2}
-\sqrt{m_p^2+q_i^2}\right)a \nn \\
&=&\left(\frac{p_i^2-q_i^2}{m_p}\right)a
+O\left(\frac{p_i^4}{m_p^3}a,\frac{q_i^4}{m_p^3}a\right).
\een
Since the terms 
$\gamma_k\pslash_0$, $\qslash_0 \gamma_k$,
$\qslash\gamma_k\pslash_0$ and $\qslash_0\gamma_k\pslash$ represent
the violation of the Lorentz symmetry 
due to the finite $m_Q a$ corrections, their coefficients
should vanish at the massless limit, 
namely $F_3^k$, $F_7^k$ $\propto$ $m_p a$ as their leading contributions.
Hence the combination of $F_3^k$, $F_7^k$ and $(p_0-q_0)a$ results in 
$O(a^2)$.
This is retained even in the case of $\lqcd\ll m_Q\sim a^{-1}$.

The relevant term for the determination of $\cb$ is the third one,
which can be extracted by setting $p=\pos\equiv (p_0=im_p, p_i=0)$ and
$q=\qos\equiv (q_0=im_p, q_i=0)$ in eq.(\ref{eq:vtx_k_offsh}):
\ben
&&\left.H_1^k+im_p (H_2^k+H_3^k)-m_p^2 H_4^k\right|_{p=\pos,q=\qos}\nn\\
&=& \frac{1}{8}{\rm Tr} 
\left[\left\{\frac{\partial}{\partial p_k}
+\frac{\partial}{\partial q_k}\right\}\Lambda_k^\nlo(\pos,\qos,m)
(\gamma_4+1)\right]\nn\\
&&-\frac{1}{8}{\rm Tr}\left[\left\{\frac{\partial}{\partial p_i}
-\frac{\partial}{\partial q_i}\right\}
\Lambda_k^\nlo(\pos,\qos,m)(\gamma_4+1)\gamma_i\gamma_k\right]^{i\ne k},
\label{eq:order_a_s}
\een
where we have used the fact that $F^k$, $G^k$ and $H^k$ are functions of
$p^2$, $q^2$ and $p\cdot q$, so that
\ben
&&\left.\frac{\partial F_j^k}{\partial p_i}\right|_{p=\pos,q=\qos}
=\left.\frac{\partial F_j^k}{\partial q_i}\right|_{p=\pos,q=\qos}=0, \\
&&\left.\frac{\partial H_l^k}{\partial p_i}\right|_{p=\pos,q=\qos}
=\left.\frac{\partial H_l^k}{\partial q_i}\right|_{p=\pos,q=\qos}=0, \\
&&\left.\frac{\partial G_l^k}{\partial p_i}\right|_{p=\pos,q=\qos}
=\left.\frac{\partial G_l^k}{\partial q_i}\right|_{p=\pos,q=\qos}=0 
\een
with $j=1,\dots,8$, $l=1,2,3,4$ and $i=1,2,3$.

We should remark that the third term in eq.(\ref{eq:vtx_k_onsh})
contains both the lattice artifact of $O(p_ka, q_ka)$ and
the physical contribution of $O(p_k/m,q_k/m)$.
The parameter $\cb$ is determined to eliminate 
the lattice artifacts of $O(p_ka, q_ka)$:
\ben
\frac{\cb^{(1)}-r_s^{(1)}}{2}&=&\left[H_1^k+i m_p (H_2^k+H_3^k)
-m_p^2 H_4^k\right]_{p=\pos,q=\qos}^\latt\nn\\
&&-Z_q^\lo\left[H_1^k+im_p (H_2^k+H_3^k)
-m_p^2 H_4^k\right]_{p=\pos,q=\qos}^\cont,
\een
where we take account of the tree level expression for the
quark-quark-gluon vertex in
eq.(\ref{eq:vtx_nrm_tree_s}) and eq.(3.51) in Ref.~\cite{akt}.

We first show the calculation of eq.(\ref{eq:order_a_s})
in the continuum theory.
The contributions of Figs.~\ref{fig:vtx_1loop} (a) and (b) are expressed as
\ben
\Lambda_{k,\cont}^{(1-a,b)}(p,q,m)
&=&\int_{-\infty}^{\infty}\frac{d^D k}{(2\pi)^D}
I_{k,\cont}^{(a,b)}(k,p,q,m)
\een
with
\ben
I_{k,\cont}^{(a)}(k,p,q,m)&=&\left(-\frac{1}{2N_c}\right)
i\gamma_\alpha\frac{1}{i(\qslash+\kslash)+\mplo}
i\gamma_k\frac{1}{i(\pslash+\kslash)+\mplo}i\gamma_\alpha
\frac{1}{k^2+\lambda^2},\\
I_{k,\cont}^{(b)}(k,p,q,m)&=&\left(-\frac{N_c}{2}\right)
i\gamma_\beta\frac{1}{i(\qslash-\kslash)+\mplo}
i\gamma_\alpha\frac{1}{k^2+\lambda^2}\frac{1}{(p-q+k)^2+\lambda^2}\\
&&\times\left[\delta_{\alpha\beta}(-p_k+q_k-2k_k)
+\delta_{k\beta}(-p_\alpha+q_\alpha+k_\alpha)
+\delta_{k\alpha}(2p_\beta-2q_\beta+k_\beta)\right],\nn
\een
where we have replaced $m_p$ with $\mplo$ in the 1-loop diagrams.
Note that Figs.~\ref{fig:vtx_1loop} (c), (d), (e) and (f) 
do not exist in the continuum.
Applying the formula of eq.(\ref{eq:order_a_s}) we obtain
\ben
(a)&:&\left(-\frac{1}{2N_c}\right)\frac{-1}{16\pi^2}\frac{1}{\mplo},\\
(b)&:&\left(-\frac{N_c}{2}\right)\frac{1}{16\pi^2}\frac{1}{\mplo}
\left[-\log\left|\frac{\mplo^2}{\lambda^2}\right|+3\right].
\een
Here it should be remarked that we find the same results 
for the time component of the vertex function 
because of the space-time symmetry in the continuum theory.

To investigate the infrared behavior of the lattice integrand 
$I_{k,\latt}^{(1)}(k,p,q,m)$ we expand it in terms of $k$.
The following terms possibly yield logarithmic divergences:
\ben
(a)&:&\left(-\frac{1}{2N_c}\right)
2\mplo^2 (\cb^\lo-r_s^\lo) J_a(k,\mplo,\lambda),\\
(b)&:&\left(-\frac{N_c}{2}\right)
\left[-\mplo(\cb^\lo+2r_s^\lo) J_b(k,\mplo,\lambda)\right.\nn\\
&&\left.+{4 Z_q^\lo}\left\{J_b(k,\mplo,\lambda)
+\mplo J_c(k,\mplo,\lambda)\right\}\right],\\
(c)&:&\left(-\frac{N_c}{4}\right)(-3)\cb^\lo J_d(k,\lambda),
\een
where
\ben
J_a(k,\mplo,\lambda)&=&\frac{1}{k^2+\lambda^2}
\frac{1}{(k^2+2i\mplo k_4)^2},\\
J_b(k,\mplo,\lambda)&=&\frac{1}{(k^2+\lambda^2)^2}
\frac{ik_4}{k^2-2i\mplo k_4},\\
J_c(k,\mplo,\lambda)&=&\frac{1}{(k^2+\lambda^2)^2}
\frac{k_i^2}{(k^2-2i\mplo k_4)^2},\\
J_d(k,\lambda)&=&\frac{1}{(k^2+\lambda^2)^2}
\een
with no sum for the index $i$.
Figures~\ref{fig:vtx_1loop} (d), (e) and (f) have no infrared divergence
as long as $m\not= 0$.
The coefficients of the logarithmic divergence 
for $J_i$ ($i=a,b,c,d$) are obtained 
by performing the integration with the cutoff $\Lambda$:
\ben
\intlat\theta(\Lambda^2-k^2) J_a(k,\mplo,\lambda)
&=&\frac{1}{16\pi^2}\frac{1}{\mplo^2}\left[
\frac{1}{2}\log\left|\frac{\Lambda^2}{\lambda^2}\right|\right.\nn\\
&&\left.-\log\left|{\frac{\Lambda+\sqrt{\Lambda^2+4 \mplo^2}}
{2\mplo}}\right|\right],\\
\intlat\theta(\Lambda^2-k^2) J_b(k,\mplo,\lambda)
&=&\frac{1}{16\pi^2}\frac{1}{\mplo}\left[
-\frac{1}{2}\log\left|\frac{\Lambda^2}{\lambda^2}\right|
+\frac{1}{2}-\frac{1}{4}\frac{\Lambda^2}{\mplo^2}
\right. \\ 
&&\left.+\frac{1}{4}\frac{\Lambda}
{\mplo^2}\sqrt{\Lambda^2+4 \mplo^2}
+\log\left|{\frac{\Lambda+\sqrt{\Lambda^2+4 \mplo^2}}
{2\mplo}}\right|\right],\nn\\
\intlat\theta(\Lambda^2-k^2) J_c(k,\mplo,\lambda)
&=&-\frac{1}{2\mplo}\intlat\theta(\Lambda^2-k^2) J_b(k,\mplo,\lambda),\\
\intlat\theta(\Lambda^2-k^2) J_d(k,\lambda)
&=&\frac{1}{16\pi^2}\left[
\log\left|\frac{\Lambda^2}{\lambda^2}\right|-1\right].
\een
 From these results we find the coefficients of the infrared divergence
in Figs.~\ref{fig:vtx_1loop} (a), (b) and (c): 
\ben
(a)&:&\left(-\frac{1}{2N_c}\right)(\cb^\lo-r_s^\lo)L,\\
(b)&:&\left(-\frac{N_c}{2}\right)\left[\frac{\cb^\lo+2r_s^\lo}{2}L
-\frac{Z_q^\lo}{\mplo}L\right],\\
(c)&:&\left(-\frac{N_c}{4}\right)(-3)\cb^\lo L,
\een
where 
\ben
L=\frac{1}{16\pi^2}\ln\left|\frac{\Lambda^2}{\lambda^2}\right|.
\een
Taking the summation the total contribution is
\ben
\left(-\frac{1}{2N_c}+\frac{N_c}{2}\right)(\cb^\lo-r_s^\lo)L
-\left(-\frac{N_c}{2}\right)\frac{Z_q^\lo}{\mplo}L.
\label{eq:sum_ir_s}
\een
If the tree level values are properly tuned as $\cb^\lo=r_s^\lo$,
we are left with $-(-N_c/2)({Z_q^\lo}/{\mplo})L$, which is exactly
the same as the infrared divergence in the continuum theory
with the correct normalization factor.

\begin{figure}[t]
\centering{
\hskip -0.0cm
\includegraphics[width=140mm,angle=0]{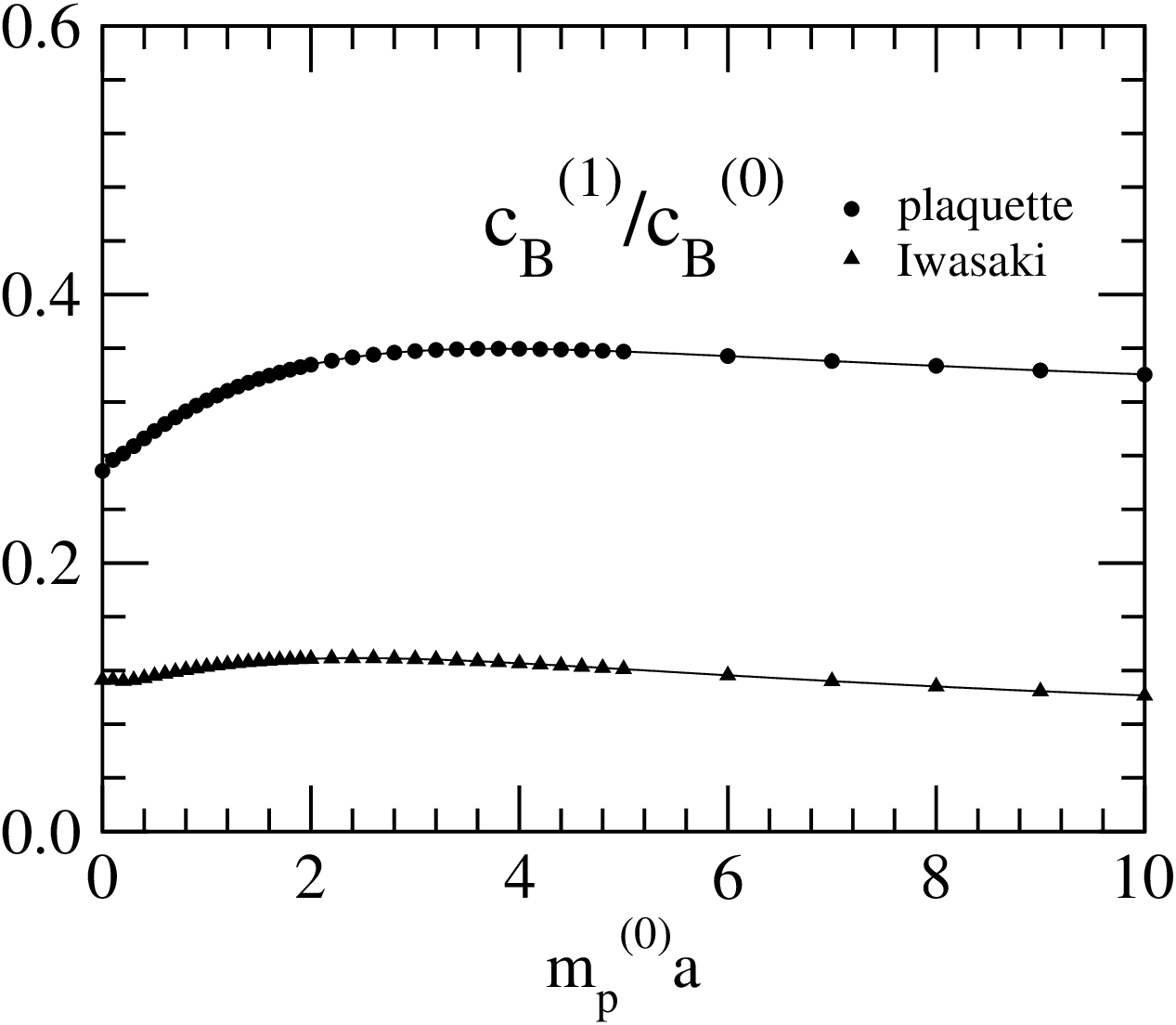}     
}
\caption{$\cb^\nlo/\cb^\lo$ as a function of $m_p^\lo$. Solid lines denote 
the interpolation of $\cb^\nlo/\cb^\lo$ with eq.(\protect{\ref{eq:fit_cb}})}
\label{fig:cb}
\vspace{8mm}
\end{figure}

Figure~\ref{fig:cb} shows
the quark mass dependences of $\cb^\nlo/\cb^\lo$ for the plaquette and the 
Iwasaki gauge actions.
We find relatively modest quark mass dependences for both cases.
The solid lines denote the fitting results of the interpolation:
\ben
\frac{\cb^\nlo}{\cb^\lo}=
\left.\frac{\cb^\nlo}{\cb^\lo}\right|_{\mplo =0}+
\frac{\sum_{i=1}^5 a_i\{\mplo\}^i}{1+\sum_{i=1}^5 b_i\{\mplo\}^i},
\label{eq:fit_cb}
\een
where the values of the parameters $a_i$ and $b_i$ ($i=1,\dots,5$) 
are given in Table~\ref{tab:fit_nu} together with $\cb^\nlo/\cb^\lo$ at
$\mplo=0$ taken from Ref.~\cite{csw_m0}.
The relative errors of this interpolation are less than a few \% 
over the range $0< \mplo\le 10$. 

We now turn to the calculation of $\ce^\nlo$.
The general form of the off-shell vertex function 
for the time component at the one-loop level is written as
\ben
\Lambda_0^{(1)}(p,q,m) 
&=&\sum_{i=a,\dots,f}\Lambda_0^{(1-i)}(p,q,m)\nn\\
&=&\sum_{i=a,\dots,f}\int_{-\pi}^{\pi}\frac{d^4 k}{(2\pi)^4}
I_0^{(i)}(k,p,q,m)\nn\\
&=&\gamma_0 F_1^0
+\gamma_0\pslash F_2^0
+\qslash \gamma_0 F_3^0
+\qslash\gamma_0\pslash F_4^0\nn\\
&&+(p_0+q_0)\left[ H_1^0+\pslash H_2^0
+\qslash H_3^0+\qslash\pslash H_4^0\right]\nn\\
&&+(p_0-q_0)\left[ G_1^0+\pslash G_2^0
+\qslash G_3^0+\qslash\pslash G_4^0\right]+O(a^2),
\label{eq:vtx_0_offsh}
\een  
where we define $\Lambda_0(p,q,m)=\Lambda_0^\lo(p,q,m)
+g^2\Lambda_0^\nlo(p,q,m)+O(g^4)$.
The coefficients $F_i^0$, $G_i^0$ and 
$H_i^0$ ($i=1,2,3,4$) are functions of $p^2$, $q^2$, $p\cdot q$ and $m$.
As in the case of $\Lambda_k$ 
the charge conjugation symmetry provides the coefficients with
the following constraint:
\ben
&&F_2^0=F_3^0,\\
&&H_2^0=H_3^0,\\
&&G_1^0=G_4^0=0,
\label{eq:g10g40}\\
&&G_2^0=-G_3^0.
\label{eq:g20g30}
\een
Sandwiching $\Lambda_0^{(1)}(p,q,m)$ by the on-shell quark states
as before,
the matrix element is reduced to
\ben
&&{\bar u}(q)\Lambda_0^{(1)}(p,q,m)u(p)\nn\\
&=&{\bar u}(q)\gamma_0 u(p)
\left\{F_1^0+i\mplo (F_2^0+F_3^0)-\mplo^2 F_4^0\right\}\nn \\
&&+(p_0+q_0){\bar u}(q) u(p)
\left\{H_1^0+i\mplo (H_2^0+H_3^0)-\mplo^2 H_4^0\right\}\nn \\
&&+(p_0-q_0){\bar u}(q) u(p)
\left\{G_1^0+i\mplo (G_2^0+G_3^0)-\mplo^2 G_4^0\right\} 
+O(a^2).
\label{eq:vtx_0_onsh}
\een
Here we replace $m_p$ with $\mplo$.
The renormalization factor is determined from the combination of
$F_1^0+i\mplo (F_2^0+F_3^0)-\mplo^2 F_4^0$, which should be
the same as 
$F_1^k+i\mplo (F_2^k+F_4^k)-\mplo^2 F_6^k$ in eq.(\ref{eq:vtx_k_onsh}).
The last term of eq.(\ref{eq:vtx_0_onsh}) vanishes 
from eqs.(\ref{eq:g10g40}) and (\ref{eq:g20g30}) as a consequence of
the charge conjugation symmetry.
We can also check $G_1^0+i\mplo (G_2^0+G_3^0)-\mplo^2 G_4^0=0$ numerically.

The coefficient $\ce$ is determined to remove the $O(a)$ contribution
of the second term in the right hand side, where 
the physical contribution of $O(p_0/m,q_0/m)$ is also included.
The second term is extracted by setting $p=\pos$ and $q=\qos$
in eq.(\ref{eq:vtx_0_offsh}) as
\ben
&&\left.H_1^0+i\mplo (H_2^0+H_3^0)-\mplo^2 H_4^0\right|_{p=\pos,q=\qos}\nn\\
&=&\frac{1}{8}{\rm Tr} 
\left[\left\{\frac{\partial}{\partial p_i}
+\frac{\partial}{\partial q_i}\right\}\Lambda_0^\nlo(\pos,\qos,m)
\gamma_i\right]\nn\\
&&+\frac{1}{i\mplo}\frac{1}{8}{\rm Tr}
\left[\Lambda_0^\nlo(\pos,\qos,m)\right]\nn\\
&&-\frac{1}{8}{\rm Tr}\left[\left\{\frac{\partial}{\partial p_i}
-\frac{\partial}{\partial q_i}\right\}\Lambda_0^\nlo(\pos,\qos,m)
\gamma_i\gamma_4\right]\nn\\
&&-i\mplo\frac{1}{4}{\rm Tr}
\left[\frac{\partial^2}{\partial p_i\partial q_j}
\Lambda_0^\nlo(\pos,\qos,m) (\gamma_4+1)\gamma_j\gamma_i\right]^{i\ne j},
\een
where we again have used the fact that $F^0$, $G^0$ and $H^0$ are 
functions of $p^2$,  $q^2$ and $p\cdot q$.
We determine the parameter $\ce$ to remove 
the $O(p_ka, q_ka)$ contributions:
\ben
\frac{\ce^{(1)}-\nu^{(1)} r_t}{2}&=&\left[H_1^0+i\mplo (H_2^0+H_3^0)
-\mplo^2 H_4^0\right]_{p=\pos,q=\qos}^\latt\nn\\
&&-Z_q^\lo\left[H_1^0+i\mplo (H_2^0+H_3^0)
-\mplo^2 H_4^0\right]_{p=\pos,q=\qos}^\cont,
\een
where we take account of the tree level expression for the
quark-quark-gluon vertex  
given in eq.(\ref{eq:vtx_nrm_tree_t}) and eq.(3.50) in Ref.~\cite{akt}.

The infrared behavior of the integrand 
$I_0^{(1)}(k,p,q,m)$ is examined by expanding it in terms of $k$.
The logarithmic divergences are attributed to 
\ben
(a)&:&\left(-\frac{1}{2N_c}\right)
\frac{J_a(k,\mplo,\lambda)}{\{{\nu^\lo}^2+r_s^\lo\sinh(\mplo)\}^2}\nn\\
&&\times\left[{\nu^\lo}^2\{4\nu^\lo\sinh(\mplo)
-3\nu^\lo\frac{{\sinh(\mplo)}^2}{\mplo}-\mplo{\nu^\lo}^2\}Z_q^\lo\right.\nn\\
&&\left.+2\nu^\lo\{\mplo\nu^\lo-\sinh(\mplo)\}
\{{\nu^\lo}^2+2r_s^\lo\sinh(\mplo)\}Z_q^\lo\right.\nn\\
&&\left.+\{-\frac{{\sinh(\mplo)}^2}{\mplo}+\mplo{\nu^\lo}^2\}
\cosh(\mplo){Z_q^\lo}^2\right.\nn\\
&&\left.+\{-\frac{{\sinh(\mplo)}^2}{\mplo}-\mplo{\nu^\lo}^2\}
r_t\sinh(\mplo){Z_q^\lo}^2\right.\nn\\
&&\left.+2\mplo\nu^\lo\ce^\lo\sinh(\mplo){Z_q^\lo}^2\right],\\
(b)&:&\left(-\frac{N_c}{2}\right)
\frac{J_b(k,\mplo,\lambda)}{{\nu^\lo}^2+r_s^\lo\sinh(\mplo)}
\left[-2\cb^\lo\sinh(\mplo)\left\{\mplo r_s^\lo+\nu^\lo \right\}\right.\nn\\
&&\left.-\frac{3}{2}\ce^\lo Z_q^\lo\left\{\nu^\lo\mplo
+\sinh(\mplo) \right\}\right.\nn\\
&&\left.+\nu^\lo\left\{-3\frac{\sinh(\mplo)}{\mplo}+\nu^\lo\right\}
\left\{\nu^\lo+r_s^\lo\mplo\right\}\right.\nn\\
&&\left.-\frac{3}{2}\nu^\lo\left\{\sinh(\mplo)-\nu^\lo\mplo \right\}
\left\{\sinh(\mplo)+r_t\cosh(\mplo) \right\}\right]\nn\\
&&+\left(-\frac{N_c}{2}\right)
\frac{J_c(k,\mplo,\lambda)}{{\nu^\lo}^2+r_s^\lo\sinh(\mplo)}
\left[-8\mplo\cb^\lo\sinh(\mplo) Z_q^\lo\right.\nn\\
&&\left.-2\nu^\lo Z_q^\lo\left\{ \sinh(\mplo)+3\mplo\nu^\lo\right\} \right], \\
(c)&:&\left(-\frac{N_c}{4}\right)(-3)\ce^\lo J_d(k,\lambda).
\een
We find no infrared divergence for Figs.~\ref{fig:vtx_1loop} (d), (e), (f)
as long as $\mplo\not= 0$.
The momentum integration with the cutoff $\Lambda$ yields
the following logarithmic divergences:
\ben
(a)&:&\left(-\frac{1}{2N_c}\right)
\frac{1}{2\mplo^2}L
\frac{1}{\{{\nu^\lo}^2+r_s^\lo\sinh(\mplo)\}^2}\nn\\
&&\times\left[{\nu^\lo}^2\{4\nu^\lo\sinh(\mplo)
-3\nu^\lo\frac{{\sinh(\mplo)}^2}{\mplo}-\mplo{\nu^\lo}^2\}Z_q^\lo\right.\nn\\
&&\left.+2\nu^\lo\{\mplo\nu^\lo-\sinh(\mplo)\}
\{{\nu^\lo}^2+2r_s^\lo\sinh(\mplo)\}Z_q^\lo\right.\nn\\
&&\left.+\{-\frac{{\sinh(\mplo)}^2}{\mplo}+\mplo{\nu^\lo}^2\}
\cosh(\mplo){Z_q^\lo}^2\right.\nn\\
&&\left.+\{-\frac{{\sinh(\mplo)}^2}{\mplo}-\mplo{\nu^\lo}^2\}
r_t\sinh(\mplo){Z_q^\lo}^2\right.\nn\\
&&\left.+2\mplo\nu^\lo\ce^\lo\sinh(\mplo){Z_q^\lo}^2\right],\\
(b)&:&\left(-\frac{N_c}{2}\right)\frac{(-1)}{2\mplo}L
\frac{1}{{\nu^\lo}^2+r_s^\lo\sinh(\mplo)}
\left[-2\cb^\lo\sinh(\mplo)\left\{\mplo r_s^\lo+\nu^\lo \right\}\right.\nn\\
&&\left.-\frac{3}{2}\ce^\lo Z_q^\lo\left\{\nu^\lo\mplo
+\sinh(\mplo) \right\}\right.\nn\\
&&\left.+\nu^\lo\left\{-3\frac{\sinh(\mplo)}{\mplo}+\nu^\lo\right\}
\left\{\nu^\lo+r_s^\lo\mplo\right\}\right.\nn\\
&&\left.-\frac{3}{2}\nu^\lo\left\{\sinh(\mplo)-\nu^\lo\mplo \right\}
\left\{\sinh(\mplo)+r_t\cosh(\mplo) \right\}\right]\nn\\
&&+\left(-\frac{N_c}{2}\right)
\frac{1}{4\mplo^2}L
\frac{1}{{\nu^\lo}^2+r_s^\lo\sinh(\mplo)}
\left[-8\mplo\cb^\lo\sinh(\mplo) Z_q^\lo\right.\nn\\
&&\left.-2\nu^\lo Z_q^\lo\left\{ \sinh(\mplo)+3\mplo\nu^\lo\right\} \right], \\
(c)&:&\left(-\frac{N_c}{4}\right)(-3)\ce^\lo L.
\een
Once we demand the tree level conditions,
\ben
&&\ce^\lo=\nu^\lo r_t,\\
&&{\nu^\lo}^2+r_s^\lo\sinh(\mplo)=\nu^\lo Z_q^\lo,\\
&&\nu^\lo=\frac{\sinh(\mplo)}{\mplo},
\een
the above expressions are reduced to be
\ben
(a)&:&0\\
(b)&:&\left(-\frac{N_c}{2}\right)\frac{1}{2\mplo}L
\left\{-2\mplo\cb^\lo+3\ce^\lo\mplo-2\nu^\lo \right\}\\
(c)&:&\left(-\frac{N_c}{4}\right)(-3)\ce^\lo L.
\een 
Finally, with the aid of another tree level condition
\ben
\cb^\lo=\frac{Z_q^\lo-\nu^\lo}{\mplo},
\een
the total contribution is found to be
\ben
-\left(-\frac{N_c}{2}\right)\frac{Z_q^\lo}{\mplo}L,
\label{eq:sum_ir_t}
\een
which is the same as that for the space component 
in eq.(\ref{eq:sum_ir_s}).

We again stress that the infrared divergences originating 
from Figs.~\ref{fig:vtx_1loop}~(a), (b), (c) 
contain both the lattice artifacts and the 
physical contributions. The former exactly cancels out if and only if
the four parameters $\nu^\lo$, $r_s^\lo$, $\cb^\lo$ and $\ce^\lo$
are properly tuned as denoted in 
eqs.(\ref{eq:nu_me}), (\ref{eq:rs_me}), (\ref{eq:ce}) and (\ref{eq:cb}). 
This is another evidence that the tree level improvement is correctly
implemented in Ref.~\cite{akt}.

\begin{figure}[b]
\centering{
\hskip -0.0cm
\includegraphics[width=140mm,angle=0]{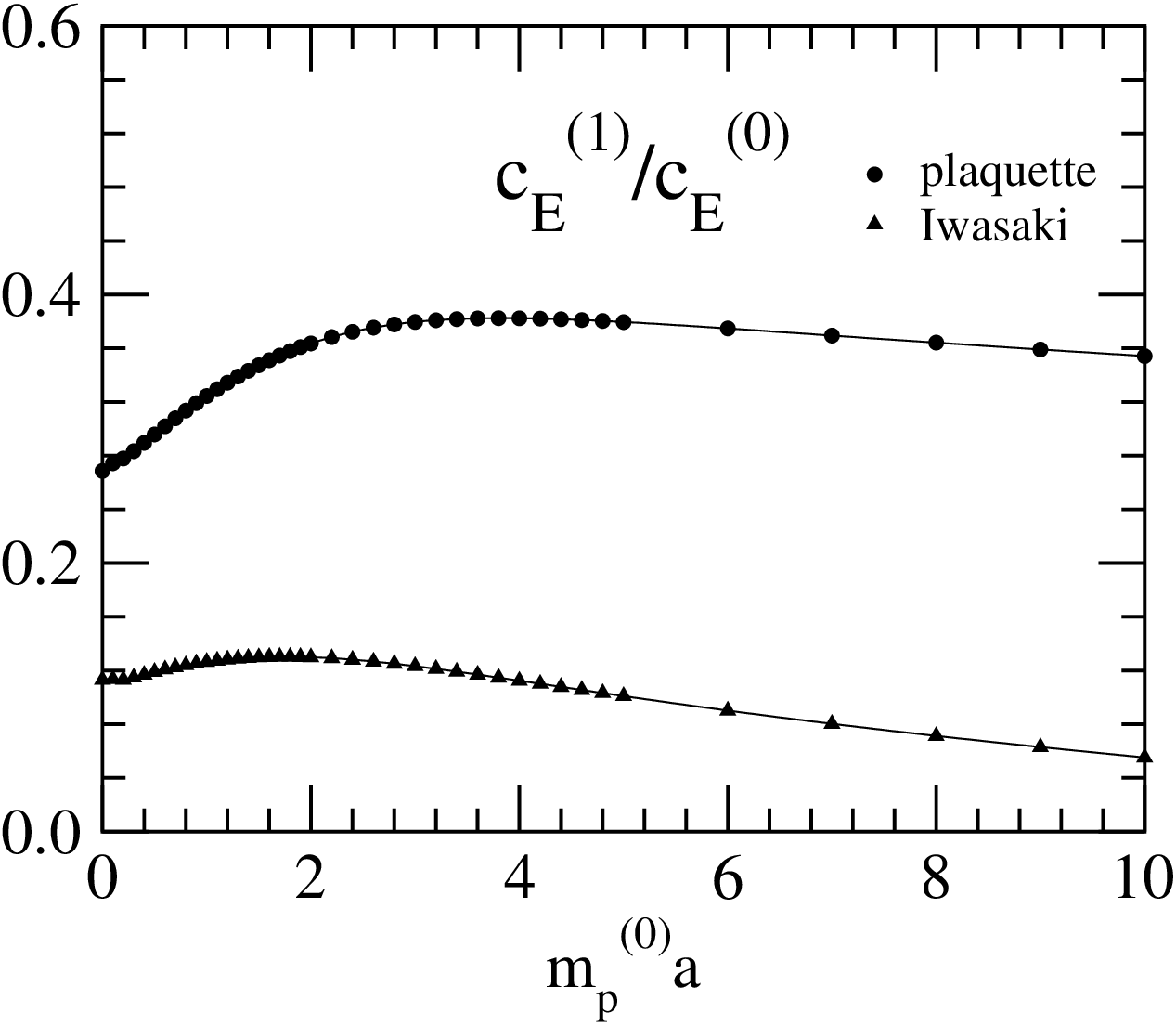}     
}
\caption{$\ce^\nlo/\ce^\lo$ as a function of $m_p^\lo$. Solid lines denote 
the interpolation of $\ce^\nlo/\ce^\lo$ with eq.(\protect{\ref{eq:fit_ce}})}
\label{fig:ce}
\vspace{8mm}
\end{figure}

In Fig.~\ref{fig:ce} we plot $\ce^\nlo/\ce^\lo$ as a function of
$\mplo$ for the plaquette and the Iwasaki gauge actions.
The fitting results of the interpolation
\ben
\frac{\ce^\nlo}{\ce^\lo}=
\left.\frac{\ce^\nlo}{\ce^\lo}\right|_{\mplo =0}+
\frac{\sum_{i=1}^5 a_i\{\mplo\}^i}{1+\sum_{i=1}^5 b_i\{\mplo\}^i}
\label{eq:fit_ce}
\een
are also shown by the solid lines.
We find relatively modest quark mass dependences similar to the $\cb$ case.
The relative errors of this interpolation are less than a few \% 
over the range $0< \mplo\le 10$.
Table~\ref{tab:fit_ce} summarizes 
the values of the parameters $a_i$ and $b_i$ ($i=1,\dots,5$)
and $\ce^\nlo/\ce^\lo$ at
$\mplo=0$ taken from Ref.~\cite{csw_m0}.

\section{Mean field improvement}

In this section we rearrange the 1-loop results in the previous sections,
using the mean-field improvement. We first replace the link variable
$U_{n,\mu}$ by $u (U_{n,\mu}/u)=u {\tilde U}_{n,\mu}$, 
where $u$ is the average of the link variable
$u=\langle U_{n,\mu}\rangle$ in some gauge fixing, or $u = P^{1/4}$
with $P$ is the average of the plaquette.
In this paper, we adopt the latter definition:
\ben
u = 1-g^2\frac{C_F}{2} T_{\rm MF}.
\een
A detailed description on the derivation of $T_\mf$ is given in Sec.~III
of Ref.~\cite{dwf_pt_rg}.

This replacement leads to the following dispersion relation.
\beqa
u \sinh (\mplomf) &=& m_0 + r_t (1-u \cosh (\mplomf))+ 3 r_s(\mplomf)(1-u)
 \nn \\
                  &=& m + r_t u (1-\cosh (\mplomf))+(1-u)3(r_s(\mplomf)-1)
\eeqa
where $m = m_0 + r_t(1-u)+3(1-u)$ and $\mplomf$ is the tree level pole mass
in the mean-field improvement. Using the relation
\beqa
\sinh (\mplo) + r_t \cosh (\mplo) &=& m + r_t ,
\eeqa
we have
\beqa
(1+u-1)(\sinh (\mplomf) + r_t \cosh (\mplomf) ) &=&
\sinh (\mplo) + r_t \cosh (\mplo) \nn \\
&+& r_t (u-1) + (1-u)3(r_s(\mplomf)-1) .
\eeqa
This leads to the relation:
\ben
\mplo& =& \mplomf + (u-1)\frac{\sinh(\mplomf)+r_t(\cosh (\mplomf)-1)
+3(r_s(\mplomf)-1)}{\cosh (\mplomf)+ r_t\sinh (\mplomf)} \nn\\
&\equiv& {\mplomf} + g^2 \Delta m_p,
\een
where
\ben
\Delta m_p &=& -\frac{ C_F}{2} T_{MF}
\frac{\sinh(\mplomf)+r_t(\cosh (\mplomf)-1)+3(r_s(\mplomf)-1)}
{\cosh (\mplomf)+ r_t\sinh (\mplomf)}. 
\een
Using this, the pole mass is rewritten as
\ben
m_p &=& \mplo + g^2 m_p^\nlo = {\mplomf}+ g^2\tilde m_p^\nlo 
\een
where $\tilde m_p^\nlo = m_p^\nlo +\Delta m_p$.

With the use of $\tilde m_p^\nlo$ we apply the mean field improvement
to $Z_q$, $\nu$, $r_s$, $\ce$ and $\cb$:
\ben
Z_{q,\latt}&=& Z_{q,\latt}^\lo(\mplomf)u
\left(1+g^2\frac{Z_{q,\latt}^\nlo}{Z_{q,\latt}^\lo}
+g^2\frac{C_F}{2}T_{\rm MF}
+\frac{g^2}{Z_{q,\latt}^\lo}\frac{\p Z_{q,\latt}^\lo}{\p \mplo} {\Delta m_p} \right), \\
\nu&=&\nu^\lo(\mplomf)+g^2\nu^\nlo(\mplomf)
+g^2\frac{\partial \nu^\lo}{\partial \mplo}{\Delta m_p} ,\\
r_s&=& r_s^\lo(\mplomf)+g^2 r_s^\nlo(\mplomf)
+g^2\frac{\partial r_s^\lo}{\partial \mplo}{\Delta m_p},\\
\ce&=&\ce^\lo \frac{1}{u^3}
\left(1+g^2\frac{\ce^\nlo}{\ce^\lo}-g^2\frac{3}{2}C_FT_\mf 
+\frac{g^2}{\ce^\lo}\frac{\p \ce^\lo}{\p \mplo}{\Delta m_p} \right), \\
\cb&=&\cb^\lo \frac{1}{u^3}\left(1+g^2\frac{\cb^\nlo}{\cb^\lo}
-g^2\frac{3}{2}C_FT_\mf
+\frac{g^2}{\cb^\lo}\frac{\p \cb^\lo}{\p \mplo}{\Delta m_p} \right).
\een
One then finally replaces $u=P^{1/4}$ with the one measured by 
Monte Carlo simulation.
 
The mean-field improved $\msbar$ 
coupling $g_\msbar^2(\mu )$ at the scale $\mu$ is obtained from the lattice
bare coupling $g_0^2$ with the use of the following relation:
\ben
\dfrac{1}{g_{\overline{\rm MS}}^2(\mu )}
&=& \dfrac{P}{g^2_0} + d_g + c_p +\dfrac{22}{16\pi^2} \log (\mu a)
+N_f\left(d_f -\dfrac{4}{48\pi^2} \log (\mu a)\right) .
\label{eq:g2_plaq}
\een
For the improved gauge action 
one may use an alternative formula\cite{cppacs}
\ben
\dfrac{1}{g_{\overline{\rm MS}}^2(\mu )}
&=& \dfrac{c_0 P + 8 c_1 R1+ 16c_2 R2 +8 c_3 R3}{g^2_0} \nn \\
& &+ d_g + (c_0\cdot c_p + 8 c_1\cdot c_{R1}+16 c_2\cdot c_{R2}+
8 c_3\cdot c_{R3})
+\dfrac{22}{16\pi^2} \log (\mu a) \nn\\
&&+N_f\left(d_f -\dfrac{4}{48\pi^2} \log (\mu a)\right),
\label{eq:g2_rg}
\een
where
\ben
P  &=& \frac{1}{3}{\rm Tr} U_{pl}    =1 - c_{p} g_0^2 +O(g_0^4),\\ 
R1 &=& \frac{1}{3}{\rm Tr} U_{rtg}    =1 - c_{R1} g_0^2 +O(g_0^4),\\
R2 &=& \frac{1}{3}{\rm Tr} U_{chr}        =1 - c_{R2} g_0^2 +O(g_0^4),\\
R3 &=& \frac{1}{3}{\rm Tr} U_{plg}=1 - c_{R3} g_0^2 +O(g_0^4),
\een
and the measured values are employed for $P$, $R1$, $R2$ and $R3$. 
The values of $c_{p}$, $c_{R1}$, $c_{R2}$ and $c_{R3}$ for various 
gauge actions are listed in Table~XVI of Ref.~\cite{dwf_pt_rg}.


\section{Conclusion}

In this paper we determine the $O(a)$ improvement coefficients, 
$\nu$, $r_s$, $\cb$ and $\ce$ 
in the relativistic heavy quark action up to the one-loop order
for the various improved gauge actions. As byproducts we also calculate
the quark wave function $Z_q$ and the pole mass $m_p$.
While $\nu$, $r_s$, $Z_q$ and $m_p$ are determined from the quark propagator,
we use the on-shell quark-quark scattering amplitude for $\cb$ and $\ce$.
The $m_Q a$ dependences are examined by making the perturbative calculations
done in the $m_Q a$ dependent way:
As for the results of $\nu^\nlo/\nu^\lo$, $r_s^\nlo/r_s^\lo$, 
$Z_q^\nlo/Z_q^\lo$ and $m_p^\nlo$ 
we observe the strong $m_Qa$ dependence for $m_Q a\simlt 1$,
while the dependence becomes much milder beyond $m_Q a\sim 1$.
On the other hand, $\cb^\nlo/\cb^\lo$ and $\ce^\nlo/\ce^\lo$  show
relatively mild $m_Q a$ dependences for $0< m_Q a\le 10$.  
Employing the conventional perturbative method with the fictitious gluon mass
to regularize the infrared divergence we show that
the parameters $\nu$, $r_s$, $\cb$ and $\ce$ in the action are determined
free from the infrared divergences.
This is achieved if and only if the tree level values for 
$\nu$, $r_s$, $\cb$ and $\ce$ are properly adjusted as presented
in Ref.~\cite{akt}.
For later convenience we give a detailed description 
about how to apply the mean field improvement to our results. 
We are now trying a numerical test of this formulation with the 
mean field improved parameters employing the heavy-heavy and heavy-light
meson systems.

\section*{Acknowledgments}
This work is supported in part by the Grants-in-Aid for
Scientific Research from the Ministry of Education, 
Culture, Sports, Science and Technology.
(Nos. 13135204, 14046202, 15204015, 15540251, 15740165.)

\newpage

\newcommand{\J}[4]{{ #1} {\bf #2} (#3) #4}
\newcommand{\MPL}{Mod.~Phys.~Lett.}
\newcommand{\IJMP}{Int.~J.~Mod.~Phys.}
\newcommand{\NP}{Nucl.~Phys.}
\newcommand{\PL}{Phys.~Lett.}
\newcommand{\PR}{Phys.~Rev.}
\newcommand{\PRL}{Phys.~Rev.~Lett.}
\newcommand{\AP}{Ann.~Phys.}
\newcommand{\CMP}{Commun.~Math.~Phys.}
\newcommand{\PTP}{Prog.~Theor.~Phys.}
\newcommand{\Suppl}{Prog. Theor. Phys. Suppl.}
\bibliography{basename of .bib file}

\newpage

\begin{table*}[htb]
\caption{Values of parameters $a_i$ and $b_i$ ($i=1,2,3$) in the interpolation 
of $\mplo$ with eq.(\protect{\ref{eq:fit_mp}}) 
for the various gauge actions.}
\label{tab:fit_mp}
\newcommand{\cc}[1]{\multicolumn{1}{c}{#1}}
\begin{tabular}{lllllll}
\hline
gauge action   & $a_1$ & $a_2$ & $a_3$ & $b_1$ & $b_2$ & $b_3$  \\
\hline
plaquette      & 0.44498 & 1.1694 & 0.20262 & 5.1026 & 3.4713 & 1.3421  \\
Iwasaki        & 0.33282 & 1.2533 & 0.24384 & 7.7008 & 7.4235 & 2.4353  \\
Symanzik       & 0.39766 & 1.2161 & 0.22280 & 5.9914 & 4.6478 & 1.6915 \\
Iwasaki'       & 0.33819 & 1.2549 & 0.23934 & 7.5388 & 7.1244 & 2.3482 \\
Wilson         & 0.32642 & 1.2626 & 0.23357 & 7.9398 & 7.8574 & 2.5026 \\
DBW2           & 0.25898 & 1.1989 & 0.24076 & 10.889 & 14.868 & 4.1077 \\
\hline
\end{tabular}
\end{table*}

\clearpage

\begin{sidetable}[htb]
\caption{Values of parameters $a_i$ and $b_i$ ($i=1,\dots,4$) 
in the interpolation of $\Delta_q^\nlo$ with eq.(\protect{\ref{eq:fit_wf}}) 
for the various gauge actions. The values of $\Delta_q^\nlo$ at $\mplo=0$
are taken from Ref.~\protect{\cite{gmass}} except for the DBW2 gauge action.}
\label{tab:fit_wf}
\newcommand{\cc}[1]{\multicolumn{1}{c}{#1}}
\begin{tabular}{llllllllll}
\hline
gauge action   & $\Delta_q^\nlo(\mplo=0)$ 
& $a_1$ & $a_2$ & $a_3$ & $a_4$ & $b_1$ & $b_2$ & $b_3$ & $b_4$  \\
\hline
plaquette      & $-$0.07773 & 0.18777 & 3.1560 & $-$0.15124 & 0.090311 & 15.929 & 11.134 & 0.83734 & 0.44445\\
Iwasaki        & $-$0.01478 & 0.088032 & 4.5405 & $-$0.59509 & 0.12110 & 48.391 & 27.665 & $-$1.7715 & 0.83723\\
Symanzik       & $-$0.05044 & 0.13224 & 4.0242 & $-$0.51558 & 0.20496 & 25.697 & 17.632 & $-$0.67533 & 1.3005\\
Iwasaki'       & $-$0.01739 & 0.093314 & 2.8033 & $-$0.37819 & 0.078356 & 28.654 & 16.445 & $-$1.1042 & 0.53359\\
Wilson         & $-$0.01068 & 0.082214 & 3.4543 & $-$0.40922 & 0.085302 & 39.349 & 21.267 & $-$1.0475 & 0.56718 \\
DBW2           & $+$0.02029 & 0.036401 & 0.33748 & 0.19697 & 0.027077 & 9.0916 & 6.0990 & 3.1026 & 0.11858\\
\hline
\end{tabular}
\end{sidetable}

\clearpage

\begin{sidetable}[htb]
\caption{Values of parameters $a_i$ and $b_i$ ($i=1,\dots,5$) 
in the interpolation 
of $\nu^\nlo$ with eq.(\protect{\ref{eq:fit_nu}}) 
for the various gauge actions.}
\label{tab:fit_nu}
\newcommand{\cc}[1]{\multicolumn{1}{c}{#1}}
\begin{tabular}{lllllllllll}
\hline
gauge action   & $a_1$ & $a_2$ & $a_3$ & $a_4$ & $a_5$ & $b_1$ & $b_2$ & $b_3$ & $b_4$ & $b_5$  \\
\hline
plaquette      & 0.0013787 & 0.23077 & 2.3586 & 0.41097 & 0.21571 & 21.776 & 23.401 & 16.822 & 1.0718 & 2.0084\\
Iwasaki        & 0.010681 & 0.10966 & 1.4455 & 0.29523 & 0.14143 & 17.745 & 29.424 & 21.954 & 1.3323 & 2.7265 \\
Symanzik       & 0.0063804 & 0.18587 & 1.9939 & 0.37153 & 0.18560 & 20.428 & 26.116 & 18.407 & 1.2431 & 2.2404\\
Iwasaki'       & 0.010248 & 0.11663 & 1.4875 & 0.30469 & 0.14531 & 17.882 & 28.950 & 21.577 & 1.3199 & 2.6761\\
Wilson         & 0.010284 & 0.10466 & 1.3679 & 0.29382 & 0.13690 & 16.951 & 28.693 & 22.473 & 1.2577 & 2.7941\\
DBW2           & 0.011730 & 0.021615 & 0.77692 & 0.18009 & 0.085451 & 13.817 & 30.339 & 31.211 & 0.83789 & 3.8899\\
\hline
\end{tabular}
\end{sidetable}

\clearpage

\begin{sidetable}[htb]
\caption{Values of parameters $a_i$ and $b_i$ ($i=1,\dots,5$) 
in the interpolation 
of $r_s^\nlo$ with eq.(\protect{\ref{eq:fit_rs}}) 
for the various gauge actions.}
\label{tab:fit_rs}
\newcommand{\cc}[1]{\multicolumn{1}{c}{#1}}
\begin{tabular}{lllllllllll}
\hline
gauge action   & $a_1$ & $a_2$ & $a_3$ & $a_4$ & $a_5$ & $b_1$ & $b_2$ & $b_3$ & $b_4$ & $b_5$  \\
\hline
plaquette      & 0.048659 & 0.23362 & 1.6049 & 0.15709 & 0.068748 & 6.9070 & 18.195 & 11.853 & 0.74864 & 0.92944\\
Iwasaki        & 0.0015650 & $-$0.14504 & 1.3871 & 0.012363 & 0.044884 & 12.493 & 27.761 & 31.031 & $-$2.4872 & 2.0163\\
Symanzik       & 0.024680 & 0.14420 & 0.92264 & 0.11171 & 0.044813 & 6.8434 & 12.778 & 12.079 & 0.010953 & 0.97640\\
Iwasaki'       & 0.012139 & $-$0.18772 & 2.3352 & 0.037267 & 0.080842 & 20.355 & 42.341 & 48.503 & $-$3.5202 & 3.1917\\
Wilson         & 0.015805 & $-$0.14494 & 1.9526 & $-$0.034214 & 0.066725 & 17.479 & 36.192 & 41.823 & $-$3.7429 & 2.5913\\
DBW2           & $-$0.038636 & $-$0.21232 & 0.44240 & $-$0.11604 & 0.015381 & 5.3026 & 20.770 & 22.879 & $-$9.2470 & 1.6088\\
\hline
\end{tabular}
\end{sidetable}

\clearpage

\begin{sidetable}[htb]
\caption{Values of parameters $a_i$ and $b_i$ ($i=1,\dots,5$) 
in the interpolation of $\cb^\nlo$ with eq.(\protect{\ref{eq:fit_cb}}) 
for the various gauge actions. The values of $\cb^\nlo/\cb^\lo$ at $\mplo=0$
are taken from Ref.~\protect{\cite{csw_m0}}.}
\label{tab:fit_cb}
\newcommand{\cc}[1]{\multicolumn{1}{c}{#1}}
\begin{tabular}{llllllllllll}
\hline
gauge action   & $\cb^\nlo/\cb^\lo(\mplo=0)$ 
& $a_1$ & $a_2$ & $a_3$ & $a_4$ & $a_5$ & $b_1$ & $b_2$ & $b_3$ & $b_4$ & $b_5$ \\
\hline
plaquette      & 0.26858825 & 0.11031 & $-$0.45775 & 1.4585 & 3.5629 & 0.19832 & $-$1.2454 & 6.9947 & 72.053 & 7.4057 & 6.4400 \\
Iwasaki        & 0.11300591 & 0.031328 & $-$0.25330 & $-$0.97944 & 4.9518 & $-$0.70873 & 11.329 & $-$16.089 & 275.53 & 27.127 & 13.769\\
Symanzik       & 0.19624449 & 0.070876 & $-$0.35318 & 0.75783 & 2.6525 & $-$0.034442 & $-$2.6878 & 6.1987 & 70.755 & 8.3789 & 5.2277\\
Iwasaki'       & 0.12036501 & 0.025289 & $-$0.18878 & 0.14872 & 1.3433 & $-$0.17271 & $-$0.73798 & 9.9278 & 71.365 & 8.0592 & 3.6373\\
Wilson         & 0.10983411 & 0.021005 & $-$0.19478 & 0.16783 & 0.93978 & $-$0.15323 & $-$0.74965 & 11.415 & 64.418 & 7.0567 & 2.8708\\
DBW2           & 0.04243181 & 8.7741 &  $-$14.110 & $-$19.180 & 1.0685 & $-$5.2296 & $-$516.57 & 1177.1 & 76.371 & 472.69 & 37.314\\
\hline
\end{tabular}
\end{sidetable}

\clearpage

\begin{sidetable}[htb]
\caption{Values of parameters $a_i$ and $b_i$ ($i=1,\dots,5$) in the interpolation 
of $\ce^\nlo$ with eq.(\protect{\ref{eq:fit_ce}}) 
for the various gauge actions. The values of $\ce^\nlo/\ce^\lo$ at $\mplo=0$
are taken from Ref.~\protect{\cite{csw_m0}}.}
\label{tab:fit_ce}
\newcommand{\cc}[1]{\multicolumn{1}{c}{#1}}
\begin{tabular}{llllllllllll}
\hline
gauge action   & $\ce^\nlo/\ce^\lo(\mplo=0)$ 
& $a_1$ & $a_2$ & $a_3$ & $a_4$ & $a_5$ & $b_1$ & $b_2$ & $b_3$ & $b_4$ & $b_5$ \\
\hline
plaquette      & 0.26858825 & 0.061511 & $-$0.52275 & 1.5301 & 0.087055 & $-$0.009142 & $-$8.4208 & 27.507 & $-$2.5724 & 3.1868 & $-$0.15206 \\
Iwasaki        & 0.11300591 & 0.014754 & $-$0.16103 & 0.47732 & $-$0.093278 & $-$0.0044554 & $-$5.8133 & 21.985 & $-$1.4805 & 1.7759 & $-$0.025592\\
Symanzik       & 0.19624449 & 0.046299 & $-$0.38345 & 1.0557 & $-$0.013178 & $-$0.0040176 & $-$7.9568 & 23.808 & $-$1.5669 & 2.4568 & $-$0.058986\\
Iwasaki'       & 0.12036501 & $-$1.0186 & 10.033 & $-$0.22227 & $-$0.57195 & 0.048774 & 533.27 & $-$178.18 & 147.82 & $-$22.688 & 0.92777\\
Wilson         & 0.10983411 & 21.659 & $-$158.46 & 271.84 & $-$50.696 & 1.1180 & $-$3836.5 & 8792.4 & $-$268.31 & 644.04 & $-$28.188\\
DBW2           & 0.04243181 & $-$0.084945 & $-$0.49122 & $-$0.61820 & 0.13749 & $-$0.069162 & $-$1.2883 & 45.147 & $-$10.501 & 2.9737 & 0.14473\\
\hline
\end{tabular}
\end{sidetable}

\clearpage

\newpage 

\begin{figure}[t]
\begin{center}

\begin{picture}(200,150)(0,0)
\ArrowLine(50,50)(100,100)
\Text(80,65)[l]{$p$}
\ArrowLine(100,100)(50,150)
\Text(80,135)[l]{$q$}
\Vertex(100,100){2}
\Gluon(100,100)(171,100){5}{5}
\Text(196,115)[c]{$k_1$, $\mu$, $A$}
\LongArrow(171,115)(151,115)
\Text(110,30)[c]{(a)}
\end{picture}

\begin{picture}(200,150)(0,0)
\ArrowLine(50,50)(100,100)
\Text(80,65)[l]{$p$}
\ArrowLine(100,100)(50,150)
\Text(80,135)[l]{$q$}
\Vertex(100,100){2}
\Gluon(100,100)(150,150){5}{5}
\Gluon(100,100)(150,50){-5}{5}
\LongArrow(150,135)(136,121)
\Text(175,145)[c]{$k_2$, $\nu$, $B$}
\LongArrow(150,65)(136,79)
\Text(175,60)[c]{$k_1$, $\mu$, $A$}
\Text(110,30)[c]{(b)}
\end{picture}

\begin{picture}(200,150)(0,0)
\ArrowLine(50,50)(100,100)
\Text(80,65)[l]{$p$}
\ArrowLine(100,100)(50,150)
\Text(80,135)[l]{$q$}
\Vertex(100,100){2}
\Gluon(100,100)(150,150){5}{5}
\Gluon(100,100)(171,100){5}{5}
\Gluon(100,100)(150,50){-5}{5}
\LongArrow(150,135)(136,121)
\Text(175,145)[c]{$k_3$, $\tau$, $C$}
\LongArrow(171,115)(151,115)
\Text(196,115)[c]{$k_2$, $\nu$, $B$}
\LongArrow(150,65)(136,79)
\Text(175,60)[c]{$k_1$, $\mu$, $A$}
\Text(110,30)[c]{(c)}
\end{picture}
\vspace{-8mm}

\end{center}
\caption{Momentum assignment for the quark-gluon vertices.}
\label{fig:fr_vtx_qg}
\vspace{8mm}
\end{figure}                                                                   

\newpage

\newpage

\newpage

\newpage

\clearpage
\newpage

\end{document}